# Pesticide mixtures may compromise ecosystem services provided by an egg parasitoid of soybean pest stink bugs: a comprehensive ecotoxicological risk assessment

Matheus Rakes[1,2*]; Maíra Chagas Morais[1,2]; Gabriel Rodrigues Palma[3]; Daniel Bernardi[1]; Anderson Dionei Grützmacher[1]; Leandro do Prado Ribeiro[2]

[1]Department of Plant Protection, Eliseu Maciel Faculty of Agronomy, Federal University of Pelotas (UFPel), Pelotas, Rio Grande do Sul, Brazil.

[2]Research Center for Family Agriculture, Agricultural Research and Rural Extension Company of Santa Catarina (CEPAF/EPAGRI), Chapecó, Santa Catarina, Brazil.

[3]Discipline of Physiology, School of Medicine, Trinity College Dublin, University of Dublin, Dublin, Ireland

*Correspondig author: E-mail: matheus.rakes@gmail.com (M. Rakes)

**Abstract**

Tank mixtures of crop protection products are widely used in agricultural systems, yet their integrated effects on biological control agents remain poorly understood. We evaluated the lethal, sublethal, and transgenerational effects of four synthetic insecticides, applied alone or in binary mixtures with three fungicides and two herbicides, on adults and pupae of *Telenomus podisi* (Hymenoptera: Scelionidae), an important egg parasitoid of pest stink bugs. In adult bioassays, $LC_{50}$ values ranged from 28.26 mg a.i. $L^{-1}$ for isocycloseram + lambda-cyhalothrin to 265.88 mg a.i. $L^{-1}$ for acephate. Overall, mixtures increased adverse effects relative to the individual products, reducing survival, parasitism, and longevity. The lowest parasitism rates occurred with isocycloseram + lambda-cyhalothrin combined with chlorothalonil or glyphosate SL formulation. Transgenerational effects were also detected, including reduced $F_1$ emergence and an increased proportion of males in some treatments. Following pupal exposure, several mixtures reduced $F_0$ emergence, whereas ethiprole + fenpropimorph and acephate + bixafen + prothioconazole + trifloxystrobin also reduced parasitism and longevity of the emerged adults. Integration of lethal and sublethal responses using the reduction coefficient ($E_x$) revealed increased toxicity for several mixtures and shifts in IOBC risk classification; notably, all insecticide + glyphosate mixtures reached $E_x$ = 100% after pupal exposure. These findings show that assessing pesticides individually may underestimate ecotoxicological risk to *T. podisi* and highlight the need to consider pesticide mixtures in integrated pest management programs.

**Introduction**

Over the past decades, agriculture in the Neotropics has undergone substantial changes, driven by the adoption of no-tillage systems and the intensification of sequential cropping systems. In major agricultural regions of the Brazilian Cerrado, for example, up to three crop cycles can currently be grown within a single agricultural year (Fuentes-Llanillo et al. 2021; Kassam et al. 2022; Vezzani et al. 2024). Although these changes have modernized production systems and increased their profitability, such intensification has promoted the formation of "green bridges," which provide a continuous supply of food resources and shelter and thereby facilitate the survival of arthropod pests within cropping systems (Panizzi and Lucini 2024; Queiroz et al. 2025). Among these arthropods, the Neotropical brown stink bug, *Euschistus heros* (Fabricius, 1798) (Hemiptera: Pentatomidae), is considered a key pest of soybean production systems, particularly in Brazil, Argentina, and Paraguay, as well as in the southeastern, northern, and western United States (Kistner 2017; Dellapé 2021; Ademokoya et al. 2022; Panizzi et al. 2022; Bueno et al. 2023; Panizzi et al. 2026). As soybean cultivation has expanded worldwide, *E. heros* has become the most frequently recorded species, accounting for up to 82% of the total phytophagous stink bugs occurring in soybean fields (Saldanha et al. 2024). This pest causes both quantitative and qualitative losses by feeding on vegetative and reproductive soybean structures (Corrêa-Ferreira and Panizzi 1999; Lucini and Panizzi 2018; Tessmer et al. 2022).

Management of the phytophagous stink bug complex has relied predominantly on chemical control through applications of synthetic insecticides (Pozebon et al. 2020). In soybean crops specifically, management of these arthropod pests accounts for more than 50% of the insecticide applications made in Brazilian fields, with as many as five insecticide sprays being performed during a single soybean growing season in some cases (Carnevalli et al. 2022; Ávila et al. 2024; Bueno et al. 2024; Saldanha et al. 2024). However, under Brazilian conditions and in other agricultural regions worldwide, insecticides are seldom applied alone, as tank mixtures are used by nearly all growers because of their operational advantages (Gazziero 2015; Sparks 2025). In this context, insecticide-fungicide combinations account for more than 30% of pesticide mixtures, whereas in areas planted with glyphosate-tolerant soybean (Roundup® Ready technology), glyphosate is mixed with insecticides in more than 70% of applications (Gazziero 2015; Gandini et al. 2020). Although synthetic insecticides are extremely important for Brazilian agriculture, their indiscriminate use (without consideration of economic thresholds and/or through calendar-based applications) may result in undesirable consequences, including the rapid reduction in susceptibility of phytophagous stink bug populations to insecticidal active ingredients (Tibola et al. 2021; Steinhaus et al. 2022; Moreira et al. 2024) and the disruption of ecosystem services provided by biological control agents, whether naturally occurring or released through augmentative or inundative biological control programs (Carvalho et al. 2019; Bueno et al. 2026).

Among more sustainable pest management strategies, the adoption of augmentative biological control has increased in recent years, with the treated area expanding by more than 20% annually (van Lenteren et al. 2021). Egg parasitoids are currently regarded as the most important biological control agents of soybean pest stink bugs, with *Telenomus podisi* Ashmead (Hymenoptera: Scelionidae) being among the most abundant species in soybean fields in the United States and Brazil (Moonga et al. 2018). Furthermore, owing to its broad adaptability to climatic variation (Braz et al. 2021), the extensive knowledge available on the bioecology of *T. podisi* (Parra et al. 2023), and advances in release techniques, including the release of pupae or pre-fed adult parasitoids using unmanned aerial vehicles (Weber et al. 2022; Ramos et al. 2024; Roswadoski et al. 2024; Ribeiro et al. 2024), this parasitoid has the potential to be released across more than 38 million hectares in Brazil alone (Parra 2019).

Given that chemical control remains a practical reality and biological control is increasingly necessary for the development of more sustainable cropping systems, a fundamental step toward establishing effective integrated pest management (IPM) programs is understanding the compatibility of these control strategies within a realistic crop-management context (Bueno et al. 2023). Therefore, studies assessing pesticide toxicity to natural enemies are essential for integrating chemical and biological control, whether through augmentative or conservation biological control approaches (Rakes et al. 2021; Bueno et al. 2026). To date, however, the available scientific evidence has largely been based on evaluating the compatibility of individual pesticides applied alone with natural enemies, including *T. podisi* [see, for example, Stecca et al. (2018); Zantedeschi et al. (2018a,b); Pazini et al. (2019); Silva et al. (2022); Santos et al. (2022)]. This approach does not fully reflect actual field conditions because the use of tank mixtures may alter the physicochemical properties of the combined xenobiotics and, consequently, substantially modify their toxicity to beneficial arthropods (Gandini et al. 2020).

In this study, our primary hypothesis was that binary mixtures of insecticides with fungicides or herbicides are incompatible with *T. podisi* because of their lethal effects on the pupal and adult stages of this egg parasitoid. As a secondary hypothesis, we predicted that these mixtures would also induce sublethal and transgenerational effects, impairing key biological traits of this biological control agent and potentially reducing its effectiveness against soybean pest stink bugs. To test these hypotheses, we assessed lethal toxicity following adult exposure and adult emergence following pupal exposure, as well as sublethal and transgenerational effects on longevity, parasitism, progeny emergence, and sex ratio in the $F_0$ and $F_1$ generations. Four insecticide formulations were evaluated either individually or in binary mixtures with three fungicides and two herbicides commonly used in soybean production to determine their effects on the egg parasitoid *T. podisi*.

## Materials and Methods

### Insects

The eggs used in the bioassays were obtained from a laboratory colony of *E. heros* established from stink bugs collected in soybean fields in Chapecó, Santa Catarina, Brazil (27°05′19″ S, 52°38′13″ W). Colony maintenance procedures followed those described by Silva et al. (2011).

The *T. podisi* colony was established from individuals obtained from Nitro BioControle (Sertãozinho, São Paulo, Brazil). The egg parasitoid was reared according to the procedures described by Peres and Corrêa-Ferreira (2004), using *E. heros* eggs as hosts. Both stink bug and parasitoid colonies were maintained in controlled-environment rooms at 25 ± 2 °C, 65 ± 10% relative humidity (RH), and a 14:10 h (L:D) photoperiod.

### Pesticides

Commercial formulations of four insecticides belonging to different chemical groups, three fungicides, and two glyphosate-based herbicides were selected for the bioassays because of their widespread use in soybean production in Brazil (Agrofit 2026). Additional information on the pesticides and commercial formulations used is provided in Table 1.

**Bioassays**

All bioassays were conducted under controlled conditions in BOD-type climate-controlled chambers (Eletrolab Indústria e Comércio de Equipamentos para Laboratório Ltda., São Paulo, SP, Brazil) maintained at 25 ± 0.5 °C, 70 ± 5% RH, a 14:10 h (L:D) photoperiod, and a light intensity of 8,380 lx. A completely randomized design was used in all bioassays. Treatments consisted of the pesticides applied individually and binary mixtures of each insecticide with each fungicide or herbicide, in addition to a negative control, totaling 30 treatments.

Two types of bioassays were conducted to assess the effects of binary mixtures of insecticides with fungicides or herbicides on the egg parasitoid *T. podisi*. The first bioassay assessed mortality (lethal effect), parasitism rate, and longevity of the $F_0$ generation, as well as adult emergence and sex ratio of the $F_1$ generation, following exposure of adult parasitoids to the treatments. Because of the high susceptibility of the adult stage, the median lethal concentrations ($LC_{50}$) of the tested insecticides were initially estimated and subsequently used in this bioassay. The second bioassay consisted of applying the same treatments to parasitized *E. heros* eggs when *T. podisi* had reached the pupal stage. Assessments included adult emergence, parasitism, and longevity of surviving $F_0$ parasitoids, as well as adult emergence and sex ratio of the $F_1$ generation. For pupal exposure, the highest label-recommended rate of each pesticide was used. The same rate criterion was maintained even for products that did not affect parasitoid emergence at this concentration to ensure methodological standardization among treatments.

**Toxicity to *T. podisi* adults**

**Determination of the median lethal concentration ($LC_{50}$) of synthetic insecticides**

The $LC_{50}$ values of acephate, bifenthrin + carbosulfan, ethiprole, and isocycloseram + lambda-cyhalothrin were determined in preliminary bioassays based on the concentrations of active ingredients specified on the labels of the respective commercial formulations. These assays were conducted in two stages. The first stage was performed to establish the concentration range of each insecticide capable of causing 5–95% mortality. For this purpose, each insecticide was diluted in distilled water to obtain a stock solution containing 1,000 µg a.i. $mL^{-1}$, from which serial 1:10 dilutions were prepared. The second stage consisted of definitive concentration–response bioassays. Based on the concentration ranges established in the first stage, approximately seven concentrations of each insecticide were prepared by diluting the stock solution with distilled water.

Forty-eight-hour-old *T. podisi* adults were anesthetized with carbon dioxide ($CO_2$) and placed in a Petri dish (90 mm diameter). A 500-µL aliquot of each treatment solution was then applied using an airbrush (OVD Importadora e Distribuidora Ltda., Novo Hamburgo, RS, Brazil) equipped with a 0.3-mm spray nozzle and operated at 20 psi. The applicator was calibrated to deliver a spray volume equivalent to 150 L $ha^{-1}$. After application, groups of 10 parasitoids were transferred to flat-bottom glass tubes (25 mm diameter × 95 mm height), with the upper opening covered with *voile* fabric. Ten replicates were used for each concentration. The negative control consisted of distilled water only. A droplet of pure honey was provided daily on the *voile* as a food source for the parasitoids. Mortality was assessed 72 h after treatment. Individuals that showed no movement after gentle stimulation with a fine-bristled brush were considered dead.

**Lethal effects ($F_0$ generation)**

Adult mortality of *T. podisi* was assessed following exposure to the $LC_{50}$ of acephate, bifenthrin + carbosulfan, ethiprole, and isocycloseram + lambda-cyhalothrin, applied individually or in binary mixtures with the fungicides bixafen + prothioconazole + trifloxystrobin, fenpropimorph, and chlorothalonil and with glyphosate herbicides formulated as soluble concentrate (SL) or water-dispersible granules (WG). Application, parasitoid handling, and mortality assessment procedures were the same as those described for $LC_{50}$ determination. The negative control consisted of distilled water only.

Five replicates were used per treatment, with each replicate consisting of 10 parasitoids that were 48 h old at the time of exposure. A droplet of pure honey was provided daily on the *voile* as a food source. Mortality was assessed 72 h after treatment, and individuals showing no movement after gentle stimulation with a fine-bristled brush were considered dead.

**Sublethal effects ($F_0$ and $F_1$ generations)**

Among the parasitoids surviving 72 h after treatment, five females were randomly selected from each treatment to assess parasitism. Females were individually placed in flat-bottom glass tubes (25 mm diameter × 95 mm height), with the upper opening covered with *voile* fabric and secured with an elastic band. Each experimental unit received a sky-blue cardstock strip containing 35 fresh *E. heros* eggs (~12 h old), which were exposed to a female parasitoid for 24 h.

After this period, the egg cards were removed and individually transferred to new glass tubes to determine the number of eggs parasitized by surviving $F_0$ females. Eggs exhibiting the characteristic black coloration associated with parasitoid development were considered parasitized. Transgenerational effects were subsequently assessed by recording the number of $F_1$ adults emerging from the parasitized eggs and the proportion of males and females in the $F_1$ progeny.

Longevity of $F_0$ parasitoids surviving treatment exposure was also evaluated using experimental units identical to those described above. Mortality was recorded daily until all parasitoids had died, and droplets of pure honey were provided *ad libitum* as a food source. Each individual parasitoid constituted one replicate, with 15 replicates used per treatment.

## Toxicity to *T. podisi* pupae

### Lethal effects ($F_0$ generation)

For the post-parasitism bioassays targeting the pupal stage, 2.5-cm² cards were prepared from sky-blue cardstock. Thirty-five *E. heros* eggs parasitized by *T. podisi* were attached to each card using colorless and odorless gum arabic. Each card was then placed in a flat-bottom glass tube (25 mm diameter × 95 mm height) and maintained in a BOD-type climate-controlled chamber at 25 °C for 216 h (9 days) after parasitism, corresponding to the pupal stage of *T. podisi* (Stecca et al. 2018).

Five replicates were subsequently assigned to each treatment for pesticide application. Each card was sprayed with 500 μL of the respective treatment solution using the same equipment described for the adult bioassays, calibrated to deliver a spray volume equivalent to 150 L $ha^{-1}$, according to protocols established by the International Organization for Biological and Integrated Control (IOBC) (Hassan et al. 2000; Stecca et al. 2018; Rakes et al. 2022).

After treatment, the cards containing sprayed *E. heros* eggs with *T. podisi* pupae were transferred to glass tubes identical to those described above and maintained in BOD-type climate-controlled chambers. Adult emergence of the $F_0$ generation was then determined following pupal exposure to the treatments. Only individuals that completely emerged from the host egg were considered successfully emerged.

**Sublethal effects ($F_0$ and $F_1$ generations)**

Among the parasitoids that successfully emerged following pupal exposure to the treatments, 15 randomly selected $F_0$ adults per treatment were individually maintained in flat-bottom glass tubes (25 mm diameter × 95 mm height) to assess longevity. Survival was recorded daily until the death of each parasitoid, and droplets of pure honey were provided as a food source.

In addition, five $F_0$ females per treatment were randomly selected to assess parasitism. The procedures followed those described above for assessing sublethal effects after adult exposure. Finally, transgenerational effects on the $F_1$ generation were assessed by recording the number of adults emerging from eggs parasitized by $F_0$ females and determining the proportion of males and females in the resulting progeny.

**Data analysis, reduction coefficient ($E_x$), and IOBC classification**

To estimate the $LC_{50}$ values of the tested insecticides for adult *T. podisi*, a binomial model with a complementary log-log link function (gompit model) was fitted using the PROBIT procedure in SAS statistical software, version 9.2 (SAS Institute 2011).

The percentages of mortality of adults and of F1 emergence after adult exposure were analyzed by fitting binomial generalized linear models (GLMs) (Nelder and Wedderburn 1972; McCullagh and Nelder 1989; Demétrio et al. 2014), including the effect of treatments in the linear predictor (logit link). The percentages of parasitism and of $F_1$ females after adult exposure, and the percentages of F0 emergence, F1 emergence and F1 females after pupal exposure, were analyzed by fitting beta-binomial GLMs, including the effect of treatments in the linear predictor (logit link).

The percentage of parasitism after pupal exposure was analyzed by fitting a beta-binomial generalized additive model for location, scale, and shape (GAMLSS) (Stasinopoulos et al. 2017), including the effect of treatments in the linear predictors for the mean (logit link) and dispersion (log link). Treatments with no variation among replicates (e.g., 100% mortality or 0% emergence in all replicates) were excluded from the models, as were treatments that could not be evaluated because of high mortality.

The longevity of the parasitoids surviving adult exposure and emerging after pupal exposure was analyzed by fitting negative binomial type II GLMs, for which the variance is a linear function of the mean, including the effect of treatments in the linear predictor (log link). The beta-binomial and negative binomial type II GLMs were fitted using the gamlss package (Stasinopoulos et al. 2017), modelling only the mean.

The significance of the treatment effects for the GAMLSS was assessed through likelihood ratio (LR) tests for nested models. When significant, pairwise differences between treatments were assessed through LR tests. Goodness-of-fit was assessed through *worm plots* for GAMLSS (van Buuren and Fredriks 2001; Stasinopoulos et al. 2017), through half-normal plots with a simulated envelope for GLMs (Moral et al. 2017), and, for Cox proportional-hazards models, by visual analysis of the martingale residuals (Therneau and Grambsch 2000). All analyses were carried out using R (R Core Team 2026).

The time until death of the parasitoids was analyzed by fitting Cox proportional-hazards models for each insecticide, including the effect of treatments (the insecticide applied alone and in binary mixtures, the fungicides, the herbicides, and the negative control) in the linear *predictor* (log link), and survival curves were estimated by the Kaplan–Meier method (Kaplan and Meier 1958).

The lethal and sublethal effects of the treatments on *T. podisi* were additionally integrated using a reduction coefficient (Ex), following the approach employed by Rakes et al. (2022) to combine different components of parasitoid biological performance into a single index. For variables expressed relative to the negative control, reduction was calculated as:

$$R\,(\%) = \max\,[0, (1 - \frac{T}{C}) \times 100]$$

where T and C represent the treatment and control means, respectively. Negative reduction values were set to zero.

All bioassays encompassed the $F_0$ and $F_1$ generations and differed according to the parasitoid life stage directly exposed. In the adult-exposure bioassays, $F_0$ parasitoids were directly exposed to the treatments; therefore, Ex integrated $F_0$ mortality (Emx), reduction in parasitism by $F_0$ adults (PR), reduction in emergence of the $F_1$ progeny ($ERF_1$), reduction in sex ratio (SR), and reduction in longevity (LR), as follows:

$$Ex\,(\%) = 100 \times \{1 - [(1 - \frac{Emx}{100})(1 - \frac{PR}{100})(1 - \frac{ER_{F1}}{100})(1 - \frac{SR}{100})(1 - \frac{LR}{100})]\}$$

In the pupal-exposure bioassays, the F0 generation was directly exposed at the pupal stage; therefore, the direct effect was represented by the reduction in $F_0$ emergence ($ERF_0$), replacing adult mortality. In this case, Ex integrated $ERF_0$, reduction in parasitism by emerged $F_0$ adults (PR), reduction in $F_1$ emergence ($ERF_1$), reduction in sex ratio (SR), and reduction in longevity (LR), as follows:

$$Ex\,(\%) = 100 \times \{1 - [(1 - \frac{ER_{F0}}{100})(1 - \frac{PR}{100})(1 - \frac{ER_{F1}}{100})(1 - \frac{SR}{100})(1 - \frac{LR}{100})]\}$$

Variables that could not be evaluated because no parasitoids survived or emerged were assigned a reduction of 100%. Ex values were classified according to the laboratory categories of the International Organization for Biological and Integrated Control (IOBC): class 1, harmless (Ex < 30%); class 2, slightly harmful (30% ≤ Ex < 80%); class 3, moderately harmful (80% ≤ Ex ≤ 99%); and class 4, harmful (Ex > 99%).

## Results

### Toxicity to *T. podisi* adults

*Lethal effects*

The estimated $LC_{50}$ values for adult T. podisi ranged from 28.26 to 265.88 mg a.i. $L^{-1}$. In ascending order, the $LC_{50}$ estimates were 28.26, 42.97, 68.27, and 265.88 mg a.i. $L^{-1}$ for isocycloseram + lambda-cyhalothrin, bifenthrin + carbosulfan, ethiprole, and acephate, respectively (Table 2). Although the 95% confidence intervals did not overlap for most comparisons, overlap was observed between bifenthrin + carbosulfan and ethiprole, precluding confirmation of a significant difference between these two estimates based solely on confidence interval overlap.

After 72 h of exposure, cumulative mortality of adult *T. podisi* differed significantly among treatments ($\chi^2 = 620.26$; df = 28; $p < 0.01$) (Table 3). Among the treatments evaluated, only the synthetic fungicides bixafen + prothioconazole + trifloxystrobin, fenpropimorph, and chlorothalonil, as well as the two glyphosate-based herbicide formulations (WG and SL), when applied individually, did not differ significantly from the negative control (Table 3). In addition, combining fenpropimorph with the $LC_{50}$ of acephate resulted in a significant increase in mortality of more than 40% compared with the $LC_{50}$ of acephate applied alone ($p < 0.01$) (Table 3).

Significant differences in survival probability were observed among adult *T. podisi* exposed to the $LC_{50}$ of isocycloseram + lambda-cyhalothrin (Fig. 1A; $\chi^2 = 55.54$; df = 9; $p < 0.01$), acephate (Fig. 1B; $\chi^2 = 54.90$; df = 11; $p < 0.01$), bifenthrin + carbosulfan (Fig. 1C; $\chi^2 = 46.01$; df = 11; $p < 0.01$), and ethiprole (Fig. 1D; $\chi^2 = 28.11$; df = 7; $p < 0.01$), applied either alone or in binary mixtures with the tested synthetic fungicides or herbicides (LR tests of the Cox *p*roportional-hazards models). When chlorothalonil, bixafen + prothioconazole + trifloxystrobin, or glyphosate (SL formulation) was combined with the $LC_{50}$ of isocycloseram + lambda-cyhalothrin, median survival time was 4.5 to 23 days shorter than that of the insecticide applied alone (39.0 days), although these differences were not significant ($p > 0.05$), whereas the mixture with fenpropimorph increased survival (43.5 days; $p = 0.03$) (Fig. 1A). The median survival time of adult *T. podisi* exposed to the binary mixture of acephate + bixafen + prothioconazole + trifloxystrobin (22.0 days) was significantly reduced by approximately 20 and 40 days compared with acephate alone (43.5 days; $p = 0.03$) and the negative control (62.0 days; $p < 0.01$), respectively, whereas acephate + glyphosate (WG formulation) increased survival compared with acephate alone (45.5 days; $p < 0.01$) (Fig. 1B). For bifenthrin + carbosulfan (Fig. 1C) and ethiprole (Fig. 1D), no binary mixture reduced survival compared with the insecticide applied alone; the mixtures of bifenthrin + carbosulfan with bixafen + prothioconazole + trifloxystrobin (53.0 days) and with glyphosate (WG formulation; 60.5 days) increased survival compared with the insecticide applied alone (48.0 days; $p = 0.02$). All insecticides, applied alone or in mixtures, differed significantly from the negative control, except acephate + glyphosate (WG formulation) and the mixtures of bifenthrin + carbosulfan with bixafen + prothioconazole + trifloxystrobin and with either glyphosate formulation ($p > 0.05$).

*Sublethal effects*

Significant differences in the percentage of eggs parasitized by females that survived exposure to the treatments ($F_0$ generation) were observed ($\chi^2 = 132.82$; df = 26; $p < 0.01$) (Table 3). Among the evaluated treatments, only the fungicides bixafen + prothioconazole + trifloxystrobin, fenpropimorph, and chlorothalonil did not significantly reduce the percentage of parasitism compared with the control (Table 3). In contrast, the lowest parasitism rates (<15%) were observed when insects were exposed to binary mixtures of the $LC_{50}$ of isocycloseram + lambda-cyhalothrin with chlorothalonil and with glyphosate (SL formulation). These treatments differed significantly from all the other treatments (Table 3).

Regarding sublethal effects on the $F_0$ generation, when adult parasitoids were exposed to the treatments, significant differences in the longevity of insects surviving 72 h after exposure were observed ($\chi^2 = 91.21$; df = 25; $p < 0.01$) (Fig. 2). Compared with the negative control (59.00 days), longevity was significantly reduced following exposure to bixafen + prothioconazole + trifloxystrobin (37.50 days), glyphosate (SL formulation, 36.00 days; WG formulation, 40.40 days), and ethiprole (40.90 days) applied alone; to the binary mixtures of acephate with bixafen + prothioconazole + trifloxystrobin (24.20 days), fenpropimorph (35.40 days), chlorothalonil (35.00 days), and glyphosate (SL formulation; 42.60 days); to the binary mixtures of bifenthrin + carbosulfan with fenpropimorph (43.20 days) and chlorothalonil (39.10 days); as well as to isocycloseram + lambda-cyhalothrin applied alone (29.70 days) or in binary mixtures with bixafen + prothioconazole + trifloxystrobin (20.30 days), fenpropimorph (39.80 days), and chlorothalonil (32.80 days). These treatments reduced parasitoid longevity by 15.8 to 38.7 days relative to the negative control (Fig. 2).

Transgenerational effects were detected in the progeny ($F_1$ generation) emerging from eggs parasitized by surviving $F_0$ females, with significant differences in progeny emergence among treatments ($\chi^2 = 201.14$; df = 25; $p < 0.01$) (Table 4). Among the treatments evaluated, only fenpropimorph applied individually and the binary mixture of acephate + glyphosate (SL formulation) did not significantly affect progeny emergence compared with the control, with emergence rates exceeding 78% (Table 4). In addition, significant differences were observed in the sex ratio of the emerged progeny ($\chi^2 = 82.70$; df = 25; $p < 0.01$) (Table 4). Among the pesticides tested individually, acephate, ethiprole, bixafen + prothioconazole + trifloxystrobin, chlorothalonil, and glyphosate (SL formulation) significantly increased the proportion of emerged males relative to the control (Table 4). Significant increases in the proportion of males were also observed for the binary mixtures of acephate with chlorothalonil or with either glyphosate formulation, as well as for the mixture of bifenthrin + carbosulfan with glyphosate (SL formulation), compared with the control (Table 4).

**Toxicity to the pupal stage of *T. podisi* (post-parasitism bioassay)**

*Lethal effects*

When the insecticides and their respective binary mixtures were applied at the highest label-recommended rates to parasitized *E. heros* eggs, corresponding to the pupal stage of the egg parasitoid *T. podisi*, significant differences in $F_0$ adult emergence were observed among treatments included in the statistical analysis ($\chi^2 = 180.47$; df = 14; $p < 0.01$) (Table 5). Among these treatments, only bixafen + prothioconazole + trifloxystrobin, chlorothalonil, and the binary mixture of ethiprole + bixafen + prothioconazole + trifloxystrobin did not differ significantly from the negative control. Additionally, no adult emergence was observed in 15 treatments, which were excluded from the statistical analysis because of the absence of variation among replicates (Table 5).

In addition, significant differences in the survival probability of *T. podisi* were observed when the pupal stage was exposed to acephate (Fig. 3A; $\chi^2 = 56.58$; df = 8; p < 0.01) or ethiprole (Fig. 3B; $\chi^2 = 40.01$;

df = 9; p < 0.01), either applied alone or in their respective binary mixtures (LR tests of the Cox *p*roportional-hazards models). Median survival time decreased from 31 to 6 days when bixafen + prothioconazole + trifloxystrobin was combined with acephate, compared with acephate applied alone, although this difference was not significant (p = 0.06), whereas acephate + chlorothalonil increased survival compared with acephate applied alone (36 vs. 31 days; p = 0.04) (Fig. 3A). In contrast, the binary mixture of fenpropimorph with ethiprole significantly reduced median survival time by approximately 50 days compared with ethiprole applied alone (4 vs. 55 days; p < 0.01) (Fig. 3B).

*Sublethal effects*

Significant differences were observed in the parasitism of the emerged parasitoids ($F_0$ generation) ($\chi^2$ = 50.40; df = 12; p < 0.01), as well as in its dispersion among replicates ($\chi^2$ = 69.84; df = 12; p < 0.01) (Table 5). Compared with the negative control, parasitism was significantly reduced by acephate applied alone or in binary mixtures with bixafen + prothioconazole + trifloxystrobin or chlorothalonil, and by ethiprole in binary mixtures with bixafen + prothioconazole + trifloxystrobin, chlorothalonil, or fenpropimorph, with reductions greater than 40% for acephate + bixafen + prothioconazole + trifloxystrobin and ethiprole + fen*p*ropimorph (Table 5). In addition, significant differences in adult longevity were observed among parasitoids that successfully completed emergence ($\chi^2$ = 97.80; df = 12; $p$ < 0.01). Compared with the control, significant reductions in $F_0$ parasitoid longevity were observed following exposure to both glyphosate formulations, acephate applied alone or in binary mixtures with bixafen + prothioconazole + trifloxystrobin or chlorothalonil, and ethiprole in binary mixtures with bixafen + prothioconazole + trifloxystrobin, chlorothalonil, or fenpropimorph (Fig. 4).

Regarding transgenerational effects, significant differences were observed in progeny emergence (F1 generation) ($\chi^2$ = 31.58; df = 12; $p$ < 0.01) (Table 6). Compared with the control, chlorothalonil, ethiprole, and acephate applied alone, as well as the binary mixtures of acephate + bixafen + prothioconazole + trifloxystrobin, ethiprole + chlorothalonil, and ethiprole + fenpropimorph, significantly reduced the percentage of emerged parasitoids (Table 6). Significant differences were also detected in the sex ratio of the emerged F1 progeny ($\chi^2$ = 56.92; df = 12; $p$ < 0.01) (Table 6). Acephate applied alone, acephate in binary mixtures with chlorothalonil or bixafen + prothioconazole + trifloxystrobin, and the binary mixture of ethiprole + fenpropimorph significantly increased the proportion of emerged males compared with the negative control (Table 6).

## Reduction coefficient in pupae and adults

When lethal and sublethal effects were integrated using the reduction coefficient ($E_x$), marked differences were observed between pesticides applied individually and their respective binary mixtures (Fig. 5). Following adult exposure, the fungicides and both glyphosate formulations, when applied individually, were classified as slightly harmful (IOBC Class 2), with $E_x$ values ranging from 45.8 to 75.0%. Among the insecticides applied alone, bifenthrin + carbosulfan was also classified as slightly harmful, whereas acephate, ethiprole, and isocycloseram + lambda-cyhalothrin were classified as moderately harmful (Class 3) (Fig. 5A).

Overall, combining insecticides with fungicides or herbicides increased the total effect on *T. podisi*. This pattern was particularly evident for bifenthrin + carbosulfan, which shifted from slightly harmful when applied alone to moderately harmful in all binary mixtures evaluated. For acephate, ethiprole, and isocycloseram + lambda-cyhalothrin, which were already classified as moderately harmful when applied individually, some binary mixtures shifted the classification to harmful (Class 4), namely ethiprole and isocycloseram + lambda-cyhalothrin combined with either glyphosate formulation, acephate combined with glyphosate (SL formulation) or fenpropimorph, and isocycloseram + lambda-cyhalothrin combined with chlorothalonil (Fig. 5A).

Following exposure during the pupal stage, bixafen + prothioconazole + trifloxystrobin was the only pesticide applied individually that was classified as harmless (Class 1), whereas the other fungicides, both glyphosate formulations, and ethiprole were classified as slightly harmful (Class 2) (Fig. 5B). Acephate was classified as moderately harmful, whereas bifenthrin + carbosulfan and isocycloseram + lambda-cyhalothrin were classified as harmful, both reaching $E_x$ values of 100%. Among the binary mixtures, ethiprole showed a marked increase in $E_x$ when combined with fenpropimorph or either glyphosate formulation. Notably, all insecticide mixtures containing glyphosate, regardless of formulation, reached an $E_x$ of 100% following pupal exposure (Fig. 5B).

# Discussion

The simultaneous use of different pesticides in a single management operation is commonly adopted to broaden the spectrum of pest control and provide operational advantages, including fewer field entries and, consequently, lower production costs (Gazziero 2015). However, pesticide tank mixtures may also result in undesirable effects, including antagonistic interactions that impair treatment performance and physicochemical changes that may compromise control of the target arthropod, reduce crop productivity, and increase the toxicity of the combined xenobiotics (Karkanis et al. 2018; Gandini et al. 2020; Morais et al. 2024). Ecotoxicological studies have traditionally focused on the effects of insecticides tested individually on beneficial arthropods and have often considered only a limited number of biological endpoints (Rakes et al. 2021; Schmidt-Jeffris 2023; Elhamalawy et al. 2024; Mata et al. 2024). In the present study, we assessed lethal, sublethal, and transgenerational effects of binary mixtures of insecticides with synthetic herbicides or fungicides on two developmental stages of the egg parasitoid *T. podisi*. Overall, our results showed that acute toxicity to both pupae and adults of *T. podisi* generally increased when insecticides were combined with fungicides or herbicides. Moreover, depending on the binary mixture, sublethal and transgenerational effects were detected, indicating that pesticide mixtures may impair biological traits directly associated with the ecosystem services provided by this biological control agent.

In Brazil, nearly all (>95%) synthetic insecticides registered for the management of phytophagous stink bugs in soybean belong primarily to three chemical groups: organophosphates, neonicotinoids, and pyrethroids (Ibama 2022; Agrofit 2026). However, insecticides belonging to these groups are frequently reported as highly toxic to beneficial arthropods (Serrão et al. 2022). In *T. podisi*, insecticides such as acephate, lambda-cyhalothrin + thiamethoxam, and bifenthrin + carbosulfan, when tested at their highest label-recommended rates, have caused complete mortality of exposed parasitoids (Pazini et al. 2017; Zantedeschi et al. 2018a). Because of the high toxicity of these compounds, we first estimated the $LC_{50}$ of the insecticides tested before evaluating their binary mixtures. Our results showed that, in some combinations, the addition of glyphosate formulations substantially increased adult mortality and markedly reduced survival over time. Tank mixing may modify the physicochemical behavior of pesticide molecules through interactions among active ingredients and formulation components, potentially affecting properties such as acid dissociation (pKa) and the octanol-water partition coefficient (Kow) (Zhang et al. 2018; Gandini et al. 2020). Changes in pKa may influence the degree of ionization and, consequently, compound solubility, whereas alterations in log Kow may modify molecular lipophilicity and thereby influence affinity for biological membranes and uptake by exposed organisms (Brooks et al. 1976; deBoer and Satchivi 2014; Ren et al. 2023).

In addition to acute mortality, individuals surviving pesticide exposure may experience substantial impairment of biological performance through sublethal effects (Bartling et al. 2024). In our study, although reductions in parasitism were observed after adult *T. podisi* were exposed to insecticides applied alone, these effects were generally more pronounced when insecticides were combined with the fungicides chlorothalonil, bixafen + prothioconazole + trifloxystrobin, or fenpropimorph, as well as in some mixtures containing glyphosate. Sublethal effects on parasitism may arise through several mechanisms, including impaired host-searching ability resulting from alterations in locomotory and reproductive behavior (Araújo et al. 2024) and pesticide-induced neurophysiological disturbances (Agathokleous et al. 2023), ultimately affecting fitness-related and life-table traits (Araújo et al. 2022). These mechanisms may also contribute to the reductions in longevity observed among surviving parasitoids. When pesticides were evaluated individually, Silva et al. (2022) found no significant reduction in the parasitism rate of *T. podisi* following tarsal exposure to ethiprole compared with the control. In contrast, exposure of *T. podisi* females to bifenthrin + carbosulfan resulted in a more than 30-fold reduction in parasitism relative to the control. Similarly, Zantedeschi et al. (2018a) reported no significant effects on parasitism when *T. podisi* was exposed by tarsal contact to dry residues of the fungicide trifloxystrobin + cyproconazole or the glyphosate-based herbicide Glyphosate Atanor®. Using a similar exposure method, Santos et al. (2022) found no significant reduction in the longevity of *T. podisi* exposed to trifloxystrobin + prothioconazole. Nevertheless, although the lethal and sublethal effects of individual pesticides on natural enemies have been extensively investigated, considerably less is known about how pesticide mixtures modify these responses, particularly when multiple biological endpoints are considered simultaneously.

In the present study, transgenerational effects were also assessed through $F_1$ progeny emergence and sex ratio. Our results indicated that exposure of adult *T. podisi* to binary mixtures of insecticides with herbicides or fungicides can substantially impair the biological performance of the $F_1$ generation, particularly by reducing progeny emergence relative to pesticides applied individually. Although progeny emergence was markedly affected by some binary mixtures, effects on sex ratio were generally less pronounced. An important exception was observed for the binary mixtures of acephate or bifenthrin + carbosulfan with glyphosate (SL formulation), in which males represented 54.84% and 58.39% of the emerged $F_1$ progeny, respectively, compared with 10.50% in the negative control (Table 4). To date, relatively few studies have investigated pesticide-induced transgenerational effects in natural enemies (Pazini et al. 2019; Dai et al. 2021; Morais et al. 2022; Rakes et al. 2024; Santana et al. 2025), and the mechanisms underlying such responses remain poorly understood. Parental effects may involve the transfer of parental condition to offspring through changes in investment in the subsequent generation (Bonduriansky & Crean 2018; Sánchez-Tójar et al. 2020; Castano-Sanz et al. 2022) or anticipatory parental effects, whereby parents alter offspring phenotype in response to environmental conditions (Burgess and Marshall 2014). In parasitoids, offspring sex allocation is also governed by a complex interaction of endogenous and exogenous factors, including environmentally induced physiological stress such as pesticide exposure (Desneux et al. 2007; Moiroux et al. 2014; Yadav & Borges 2024). For example, Rakes et al. (2022) detected changes in the $F_1$ sex ratio of *T. podisi* when parental parasitoids were exposed by tarsal contact to 5-day-old residues of the cyhalofop-butyl-based herbicide. Likewise, in the lepidopteran egg parasitoid *Trichogramma pretiosum* Riley (Hymenoptera: Trichogrammatidae), Costa et al. (2023) reported significant reductions in progeny emergence together with a substantial increase in the proportion of males following exposure to thiodicarb.

The impact of a pesticide may range from harmless to harmful depending on the developmental stage of the parasitoid exposed (Stecca et al. 2018; Theenoor et al. 2024). Compared with adults, immature stages of egg parasitoids are generally afforded some protection by the host egg and its surrounding layers, which can reduce direct exposure to pesticide residues. Nevertheless, susceptibility may vary substantially throughout immature development, and the pupal stage can be particularly vulnerable to pesticide exposure (Hassan et al. 2009; Rakes et al. 2021; Gill et al. 2022). This pattern was evident in our study, as exposure of parasitized *E. heros* eggs containing *T. podisi* pupae to isocycloseram + lambda-cyhalothrin, acephate, or bifenthrin + carbosulfan, as well as several of their binary mixtures with herbicides or fungicides, severely impaired parasitoid emergence. Conversely, effects on parasitism and subsequent transgenerational responses were generally less pronounced after pupal exposure than after direct exposure of adults. The toxicity observed following treatment of parasitized host eggs may depend, at least in part, on interactions between the physicochemical properties of the spray solution and the structural and permeability characteristics of the host eggshell. The insect chorion is a complex, multilayered extracellular matrix composed predominantly of structural proteins, with lipids and carbohydrates also contributing to its organization and permeability. Consequently, pesticide penetration may vary according to both eggshell properties and physicochemical characteristics of the compounds involved. However, the relationship between pesticide physicochemical properties and penetration through the host egg to an immature parasitoid remains poorly characterized, particularly for pesticide mixtures.

Importantly, integration of lethal and sublethal endpoints through the reduction coefficient ($E_x$) reinforced the greater ecotoxicological risk associated with pesticide mixtures. In several cases, combining an insecticide with a fungicide or herbicide shifted the treatment to a higher IOBC toxicity class than that assigned to the insecticide applied alone. This pattern was particularly evident for bifenthrin + carbosulfan following adult exposure and for ethiprole following pupal exposure. Moreover, all insecticide–glyphosate mixtures reached an $E_x$ of 100% after pupal exposure, despite glyphosate alone being classified as only slightly harmful. These findings highlight that selectivity classifications derived from individual pesticides may not adequately predict the overall impact of their mixtures on *T. podisi*, and further emphasize the importance of incorporating multiple biological endpoints into ecotoxicological risk assessments.

In conclusion, our findings demonstrate that considering pesticide active ingredients individually may underestimate their overall effects on *T. podisi* when these compounds are used in binary mixtures. By jointly assessing lethal, sublethal, and transgenerational responses following exposure during both the adult and pupal stages, this study provides a more comprehensive characterization of the potential ecotoxicological risks associated with pesticide mixtures commonly used in soybean production. The pronounced effects observed for several mixtures indicate that pesticide compatibility with biological control agents cannot necessarily be inferred from the selectivity profile of the individual products. These responses may compromise biological traits that are essential for the effectiveness and persistence of *T. podisi* populations and, consequently, may reduce the contribution of this parasitoid to biological control under field conditions.

**Authors' contributions**

MR: Conceptualization, Methodology, Investigation, Data Curation, Formal Analysis, Writing – Original Draft. MCM: Investigation, Data Curation, Writing – Review & Editing. GRP: Data Curation, Formal Analysis, Writing – Review & Editing. DB, ADG and LPR: Conceptualization, Resources, Supervision, Project administration, Funding acquisition, Writing – Review & Editing. All authors read and approved the final version of this manuscript.

**Declarations**

**Funding**

The authors wish to thank the Brazilian National Counsel of Technological and Scientific Development (CNPq, grant 310385/2022-9), Foundation for Research and Innovation of the State of Santa Catarina (FAPESC, grant 2023TR000220). We also thank the CNPq (process 163544/2021-2) for scholarship conceives to the first author. The opinions, findings and conclusions or recommendations expressed in this material are those of the authors and do not necessarily reflect the views of the funding agencies.

**Conflict of interest**

The authors declare no conflict of interest.

**Availability of data and material**

Not applicable

**Code availability**

Not applicable

**Ethical approval**

This study does not contain any studies with human participants or large animals performed by any of the authors. No approval of research ethics committees was required to accomplish this study because it was conducted on insects and crop plants.

**Consent to participate**

The authors declare consent to participate in this article.

**Consent for publication**

The authors declare consent for the publication of this article.

**References**

Ademokoya B, Athey K, Ruberson J (2022). Natural enemies and biological control of stink bugs (Hemiptera: Heteroptera) in North America. Insects, 13:932. https://doi.org/10.3390/insects13100932

Agathokleous E, Blande JD, Masui N, Calabrese EJ, Zhang J, Sicard P, Guedes RN, Benelli G (2023) Sublethal chemical stimulation of arthropod parasitoids and parasites of agricultural and environmental importance. Environ Res 15:116876. https://doi.org/10.1016/j.envres.2023.116876

Agrofit. Brasil. Ministério da Agricultura, Pecuária e Abastecimento. 2025. http://agrofit.agricultura.gov.br/agrofit_cons/principal_agrofit_cons. Acessed in 10 january 2025.

Araújo MB, Peglow SV, Rakes M, Escher JP, Ribeiro LP, Bernardi D, Zefa E, Grützmacher AD (2024). Oviposition behavior of the parasitoid *Telenomus remus* (Hymenoptera: Scelionidae) on eggs of *Spodoptera frugiperda* (Lepidoptera: Noctuidae). An Acad Bras Cienc 7:e20240632. https://doi.org/10.1590/0001-3765202420240632

Araújo MB, Rakes M, Pasini RA, Bernardi D, Grützmacher AD (2022). Effect of insecticides on biological attributes of the egg parasitoid *Telenomus remus* (Hymenoptera: Platygastridae). Biocontrol Sci Technol, 32:877-890. https://doi.org/10.1080/09583157.2022.2060495

Ávila CJ, Vessoni IC, Silva IF, Vieira ECS, Mariani A, Moreira SCS (2024) IPM in soybeans: how to reduce crop damage and increase profit for the farmer. Revi Agric Neotrop 11:e8214-e8214. https://doi.org/10.32404/rean.v11i2.8214

Bartling MT, Brandt A, Hollert H, Vilcinskas A (2024). Current insights into sublethal effects of pesticides on insects. Int J Mol Sci, 25:6007. https://doi.org/10.3390/ijms25116007

Bonduriansky R, Crean AJ (2018) What are parental condition-transfer effects and how can they be detected? Methods Ecol Evol 9:450-456. https://doi.org/10.1111/2041-210X.12848

Braz ÉC, Bueno ADF, Colombo FC, Queiroz AP (2021) Temperature impact on *Telenomus podisi* emergence in field releases of unprotected and encapsulated parasitoid pupae. Neotrop Entomol 50:462-469. https://doi.org/10.1007/s13744-021-00857-3

Brooks GT (1976) Penetration and distribution of insecticides. In: Wilkinson CF (ed) Insecticide biochemistry and physiology, 1976th edn. Springer US, Boston, pp 3-58.

Bueno ADF, Sutil WP, Jahnke SM, Carvalho GA, Cingolani MF, Colmenarez YC, Corniani N (2023) Biological Control as part of the soybean integrated pest management (IPM): Potential and challenges. Agronomy 13:2532. https://doi.org/10.3390/agronomy13102532

Bueno AF, Colmenarez YC, Carnevalli RA, Sutil WP (2023) Benefits and perspectives of adopting soybean-IPM: the success of a Brazilian programme. Plant Health Cases phcs20230006:1-16. https://doi.org/10.1079/planthealthcases.2023.0006

Bueno AF, Sutil WP, Hoshino A, Silva GV (2024) The importance and challenges of a precise diagnosis for IPM: seeing the problem from different perspectives within the soybean-corn production system. Entomol Commun 6:ec06011-ec06011. https://doi.org/10.37486/2675-1305.ec06011

Bueno AF, Hoback WW, Colmenarez YC, Valmorbida I, Sutil WP, Zang LS, Horikoshi RJ (2026) Advancements, challenges, and future perspectives of soybean-integrated pest management, emphasizing the adoption of biological control by the major global producers. Plants 15:366. https://doi.org/10.3390/plants15030366

Burgess SC, Marshall DJ (2014). Adaptive parental effects: the importance of estimating environmental predictability and offspring fitness appropriately. Oikos, 123:769-776. https://doi.org/10.1111/oik.01235

Carnevalli RA, Oliveira AB, Gomes EC, Possamai EJ, Silva GC, Reis EA, Roggia S, Prando AM, Lima D (2022) Resultados do manejo integrado de pragas da soja na safra 2021/2022 no Paraná. Embrapa Soja,

Londrina, Doc 448. http://www.infoteca.cnptia.embrapa.br/infoteca/handle/doc/1147914. Accessed 05 May 2025.

Carvalho GA, Grützmacher AD, Passos LC, Oliveira RL (2019). Physiological and ecological selectivity of pesticides for natural enemies of insects. In: Souza B, Vázquez L, Marucci R (Eds.) Natural enemies of insect pests in Neotropical agroecosystems. Springer, Cham, pp 469-478

Castano-Sanz V, Gomez-Mestre I, Garcia-Gonzalez F (2022) Evolutionary consequences of pesticide exposure include transgenerational plasticity and potential terminal investment transgenerational effects. Evolution, 76:2649-2668. https://doi.org/10.1111/evo.14613

Chapman RF, Simpson SJ, Douglas A. The insects structure and function, 2013. Fifth edition. Cambridge University Press, New York USA.

Corrêa-Ferreira BS, Panizzi AR (1999). Percevejos da soja e seu manejo, Circular Técnica nº 24. Embrapa-CNPSo, Londrina.

Costa MA, Farias ES, Andrade ED, Carvalho VC, Carvalho GA (2023). Lethal, sublethal and transgenerational effects of insecticides labeled for cotton on immature *Trichogramma pretiosum*. J Pest Sci, 96:119-127. https://doi.org/10.1007/s10340-022-01481-9

Dai C, Ricupero M, Wang Z, Desneux N, Biondi A, Lu Y (2021). Transgenerational effects of a neonicotinoid and a novel sulfoximine insecticide on the harlequin ladybird. Insects, 12:681. https://doi.org/10.3390/insects12080681

DeBoer GJ, Satchivi N (2014) Comparison of translocation properties of insecticides versus herbicides that leads to efficacious control of pests as specifically illustrated by isoclast™ active, a new insecticide, and arylex™ active, a new herbicide. In: Myung K, Satchivi NM, Kingston CK (eds) Retention, Uptake, and Translocation of Agrochemicals in Plants, v1171, American Chemical Society, pp 75-93.

Dellape G (2021). An update of the distribution of the stink bugs from Argentina (Hemiptera: Pentatomidae). Rev Soc Entomol Arg, 80:23-32. https://doi.org/10.25085/rsea.800103

Demétrio CGB, Hinde J, Moral RA (2014) Models for overdispersed data in entomology. In: Ferreira CP, Godoy WAC (eds) Ecological modelling applied to entomology. Springer, Cham, pp 219-259. https://doi.org/10.1007/978-3-319-06877-0_9

Desneux N, Decourtye A, Delpuech JM (2007) The sublethal effects of pesticides on beneficial arthropods. Annu Rev Entomol 52:81-106. https://doi.org/10.1146/annurev.ento.52.110405.091440

Donoughe S (2022). Insect egg morphology: evolution, development, and ecology. Curr Opin Insect Sci 50:100868. https://doi.org/10.1016/j.cois.2021.12.008

Elhamalawy O, Bakr A, Eissa F (2024) Impact of pesticides on non-target invertebrates in agricultural ecosystems. Pestic Biochem Physiol 202:105974. https://doi.org/10.1016/j.pestbp.2024.105974

Fuentes-Llanillo R, Telles TS, Junior DS, Melo TR, Friedrich T, Kassam A (2021) Expansion of no-tillage practice in conservation agriculture in Brazil. Soil Till Res 208:104877. https://doi.org/10.1016/j.still.2020.104877

Gandini EMM, Costa ESP, Santos JB, Soares MA, Barroso GM, Corrêa JM, Zanuncio JC (2020) Compatibility of pesticides and/or fertilizers in tank mixtures. J Cleaner Produc 268:122152. https://doi.org/10.1016/j.jclepro.2020.122152

Gazziero DLP (2015) Mixtures of pesticides in tank, in Brazilian farms. Planta Daninha, 33:83-92. https://doi.org/10.1590/S0100-83582015000100010.

Gill JK, Sangha KS, Shera PS, Sharma S (2022). Insecticidal toxicity to *Trichogramma chilonis* Ishii (Hymenoptera: Trichogrammatidae) and subsequent effects on parasitic efficiency and adult emergence

rate of descendant generation. Int J Trop Insect Sci, 42:3489-3498. https://doi.org/10.1007/s42690-022-00873-9

Hassan HA (2009) Efficiency of some new insecticides on physiological, histological and molecular level of cotton leafworm. Egypt Acad J Biol Sci. A, Entomol, 2:197-209. https://doi.org/10.21608/eajbsa.2009.15443

Ibama (2022) Instituto Brasileiro do Meio Ambiente e dos Recursos Naturais Renováveis. http://www.ibama.gov.br/agrotoxicos/relatorios-de-comercializacao-de-agrotoxicos. Accessed January 06 2025

Kaplan EL, Meier P (1958) Nonparametric estimation from incomplete observations. J Am Stat Assoc 53:457-481. https://doi.org/10.1080/01621459.1958.10501452

Karkanis AC, Vellios E, Grigoriou F, Gkrimpizis T, Giannouli P (2018) Evaluation of efficacy and compatibility of herbicides with fungicides in durum wheat (*Triticum durum* Desf.) under different environmental conditions: Effects on grain yield and gluten content. Not Bot Hort Agrobot Cluj-Napoca 46:601-607. https://doi.org/10.15835/nbha46211208

Kassam A, Friedrich T, Derpsch R (2022) Successful experiences and lessons from conservation agriculture worldwide. Agronomy 12:769. https://doi.org/10.3390/agronomy12040769

Kistner EJ (2017). Climate change impacts on the potential distribution and abundance of the brown marmorated stink bug (Hemiptera: Pentatomidae) with special reference to North America and Europe. Environ Entomol, 46:1212-1224. https://doi.org/10.1093/ee/nvx157

Lucini T, Panizzi AR (2018). Electropenetrography (EPG): a breakthrough tool unveiling stink bug (Pentatomidae) feeding on plants. Neotrop Entomol, 47:6-18. https://doi.org/10.1007/s13744-017-0574-3

Mata L, Knapp RA, McDougall R, Overton K, Hoffmann AA, Umina PA (2024) Acute toxicity effects of pesticides on beneficial organisms–Dispelling myths for a more sustainable use of chemicals in agricultural environments. Sci Total Environ 930:172521. https://doi.org/10.1016/j.scitotenv.2024.172521

McCullagh P, Nelder JA (1989) Generalized linear models, 2nd edn. Chapman and Hall, London

Moiroux J, Brodeur J, Boivin G (2014) Sex ratio variations with temperature in an egg parasitoid: behavioural adjustment and physiological constraint. Anim Behav 91:61-66. https://doi.org/10.1016/j.anbehav.2014.02.021

Moonga MN, Kamminga K, Davis JA (2018). Status of stink bug (Hemiptera: Pentatomidae) egg parasitoids in soybeans in Louisiana. Environ Entomol, 47:1459-1464. https://doi.org/10.1093/ee/nvy154

Morais MC, Rakes M, Balzan F, Grützmacher AD, Bernardi D, Ribeiro LDP (2024) Effects of mixtures of arrestants, dislodgers, and phagostimulants with synthetic insecticides on green-belly stink bug: walking behavior assessments, physicochemical interactions, and laboratory and field trials. J Econ Entomol 117:167-177 https://doi.org/10.1093/jee/toad215

Morais MC, Rakes M, Pasini RA, Grützmacher AD, Nava DE, Bernardi D. (2022). Toxicity and transgenerational effects of insecticides on *Trichopria anastrephae* (Hymenoptera: Diapriidae). Neotrop Entomol, 51:143-150. https://doi.org/10.1007/s13744-021-00925-8

Moral RA, Hinde J, Demétrio CGB (2017) Half-normal plots and overdispersed models in R: the hnp package. J Stat Softw 81:1-23. https://doi.org/10.18637/jss.v081.i10

Moreira RP, Palharini RB, Massoli GS, Diniz LH, Godoy DN, Warpechowski LF, Steinhaus EA, Stürmer GR, Bernardi, O (2024) Geographic and interspecific variation in susceptibility of *Euschistus heros* and *Diceraeus furcatus* (Hemiptera: Pentatomidae) to selected insecticides in southern Brazil. Crop Prot 179:106625. https://doi.org/10.1016/j.cropro.2024.106625

Nelder JA, Wedderburn RWM (1972) Generalized linear models. J R Stat Soc 135:370-384 https://doi.org/10.2307/2344614.

Panizzi AR, Lucini T (2024) Life history studies and their implications for management of stink bugs. In: Bueno AF, Panizzi AR (eds) Stink bugs (Hemiptera: Pentatomidae) research and management: Recent advances and case studies from Brazil, Europe, and USA, 1st edn. Springer Nature Switzerland, pp 41-50.

Panizzi AR, Lucini T, Aldrich JR (2022) Dynamics in pest status of phytophagous stink bugs in the Neotropics. Neotrop Entomol 51:18–31. https://doi.org/10.1007/s13744-021-00928-5

Panizzi AR, McPherson JE, Bundy CS, Esquivel JF, Pozzebon A, Mele A, Scaccini D, Musolin DL, Karpun NN, Neimorovets VV, Javahery M, Numata H, Shintani Y, Lim UT, Zang LS, Chen YM, Mensah RK, Miles MM, Walter GH, Jackai LEN, Dingha BN, Bueno AF (2026) The expanding impact of pentatomoid bugs: drivers, challenges, and innovations in Integrated Pest Management (IPM). Entomol Gen 46: 341-364. https://doi.org/10.1127/entomologia/3800

Parra JRP (2019) Controlee biológico na agricultura brasileira. Entomol Commun 1:ec01002. https://doi.org/10.37486/2675-1305.ec01002

Parra LM, Carvalho JRD, Hoback WW, Oliveira RCD (2023) Optimizing mass rearing of the egg parasitoid, *Telenomus podisi*, for control of the brown stink bug, *Euschistus heros*. Insects 14:435. https://doi.org/10.3390/insects14050435

Pazini JDB, Padilha AC, Cagliari D, Bueno FA, Rakes M, Zotti MJ, Martins JFS, Grützmacher AD (2019) Differential impacts of pesticides on *Euschistus heros* (Hem.: Pentatomidae) and its parasitoid *Telenomus podisi* (Hym.: Platygastridae). Sci Rep 9:6544. https://doi.org/10.1038/s41598-019-42975-4

Pazini JDB, Pasini RA, Seidel EJ, Rakes M, Martins JFDS, Grützmacher AD (2017) Side-effects of pesticides used in irrigated rice areas on *Telenomus podisi* Ashmead (Hymenoptera: Platygastridae). Ecotoxicol 26:782-791. https://doi.org/10.1007/s10646-017-1809-0

Peres WAA, Corrêa-Ferreira BS (2004) Methodology of mass multiplication of *Telenomus podisi* Ash. and *Trissolcus basalis* (Woll.) (Hymenoptera: Platygastridae) on eggs of *Euschistus heros* (Fab.) (Hemiptera: Pentatomidae). Neotrop Entomol 33:457-462. https://doi.org/10.1590/S1519-566X2004000400010

Pozebon H, Marques RP, Padilha G, O´Neal M, Valmorbida I, Bevilaqua JG, Tay WT, Arnemann JA. (2020). Arthropod invasions versus soybean production in Brazil: A review. J Econ Entomol, 113:1591-1608. https://doi.org/10.1093/jee/toaa108

Queiroz AP, Panizzi AR, França-Neto JDB, Bueno AF (2025) Management strategies for the Control of *Diceraeus melacanthus* (Dallas) in Soybean (Summer)–Maize (Fall/Winter) successions. Neotrop Entomol 54:1-12. https://doi.org/10.1007/s13744-024-01219-5

R Development Core Team (2021) R - A language and environment for statistical computing. Version 3.6.2. http://r-project.org

R Core Team (2026) R: A language and environment for statistical computing. R Foundation for Statistical Computing, Vienna, Austria. Version 4.6.1. https://www.R-project.org

Rakes M, Morais, MC, Ribeiro LP, Palma GR, Moral, RA, Bernardi D, Grützmacher AD (2024). Temperature-altering effects of synthetic insecticides on the parasitoid wasp *Telenomus podisi* for the biocontrol of pentatomids in soybean crops. J Crop Health, 76:1027-1038. https://doi.org/10.1007/s10343-024-01020-9

Rakes M, Pasini RA, Morais MC, Araujo MB, Pazini JB, Seidel EJ, Bernardi D, Grützmacher AD (2021) Pesticide selectivity to the parasitoid *Trichogramma pretiosum*: A pattern 10-year database and its implications for Integrated Pest Management. Ecotoxicol Environ Saf 208:111504. https://doi.org/10.1016/j.ecoenv.2020.111504

Rakes M, Pasini RA, Morais MC, Zanella R, Prestes OD, Bernardi D, Grützmacher AD (2022) Residual effects and foliar persistence of pesticides used in irrigated rice on the parasitoid *Telenomus podisi* (Hymenoptera: Platygastridae). J Pest Sci 95, 1121–1133. https://doi.org/10.1007/s10340-021-01436-6

Ramos GS, Dalbianco AB, Santos DM, Alvarez DL, Oliveira R (2024) Management of *Euschistus heros* with the release of *Telenomus podisi* in soybean in Brazil. Bio Control 69:529-537. https://doi.org/10.1007/s10526-023-10235-0

Ren Y, Li Y, Ju Y, Zhang W, Wang Y (2023) Insect cuticle and insecticide development. Arch Insect Biochem Physiol 114:e22057. https://doi.org/10.1002/arch.22057

Ribeiro LP, Frassini G, Nesi CN, Rakes M (2024) Inoculative releases of *Telenomus podisi* Ashmead (Hymenoptera: Scelionidae) for management of the stink bug complex in soybean crops. Phytoparasitica 52:78. https://doi.org/10.1007/s12600-024-01193-z

Roswadoski L, Sutil WP, Carneiro GS, Maciel RM, Coelho Jr A, Bueno AF (2024) Release strategy and egg parasitism of *Telenomus podisi* adults fed with different diets. Bio Control 198:105626. https://doi.org/10.1016/j.bioControle.2024.105626

Saldanha AV, Horikoshi R, Dourado P, Lopez-Ovejero RF, Berger GU, Martinelli S, Head GP, Moraes T, Corrêa AS, Schwertner CF (2024) The first extensive analysis of species composition and abundance of stink bugs (Hemiptera: Pentatomidae) on soybean crops in Brazil. Pest Manag Sci 80:3945-3956. https://doi.org/10.1002/ps.8097

Sánchez-Tójar A, Lagisz M, Moran NP, Nakagawa S, Noble DW, Reinhold K (2020) The jury is still out regarding the generality of adaptive ‘transgenerational’effects. Ecology letters, 23:1715-1718. https://doi.org/10.1111/ele.13479

Santana EDR, Thiesen LV, Ribeiro LP, Takahashi TA, Parra JRP, Yamamoto PT (2025). Harmful to parents, harmless to offspring: lethal and transgenerational effects of botanical and synthetic insecticides on the egg parasitoid *Trichogramma atopovirilia*. Insects, 16:493. https://doi.org/10.3390/insects16050493

Santos LV, Alvarez DL, Santos LM, Prado JC, Bueno RCO, Hoback WW (2022) Selectivity and sub-lethal effect of pesticides on the immature and adult stages of *Telenomus podisi* (Hymenoptera: Scelionidae). Int J Trop Insect Sci 42:1731-1736. https://doi.org/10.1007/s42690-021-00698-y

Sparks TC (2025). Insecticide mixtures—uses, benefits and considerations. Pest Manag Sci, 81:1137-1144. https://doi.org/10.1002/ps.7980

SAS Institute (2011). Statistical Analysis System: Getting Started with the SAS Learning, 9.2 ed. SAS Institute, Carry, NC.

Schmidt-Jeffris RA (2023). Nontarget pesticide impacts on pest natural enemies: progress and gaps in current knowledge. Curr Opin Insect Sci, 58:101056. https://doi.org/10.1016/j.cois.2023.101056

Serrão JE, Plata-Rueda A, Martínez LC, Zanuncio JC (2022). Side-effects of pesticides on non-target insects in agriculture: A mini-review. Sci Nature, 109:17. https://doi.org/10.1007/s00114-022-01788-8

Silva DM, Carvalho GA, Souza WR, Bueno ADF (2022) Toxicity of insecticides to the egg parasitoids *Telenomus podisi* and *Trissolcus teretis* (Hymenoptera: Scelionidae). Rev Bras Entomol 66:e20220035. https://doi.org/10.1590/1806-9665-RBENT-2022-0035

Silva FAC, Calizotti GS, Panizzi AR (2011) Survivorship and egg production of phytophagous pentatomids in laboratory rearing. Neotrop Entomol 40:35-38. https://doi.org/10.1590/S1519-566X2011000100005

Stasinopoulos DM, Rigby RA, Heller GZ, Voudouris V, De Bastiani F (2017) Flexible regression and smoothing: using GAMLSS in R. Chapman and Hall/CRC, Boca Raton

Stecca CS, Bueno AF, Pasini A, Silva DM, Andrade K, Zirondi Filho DM (2018) Impact of insecticides used in soybean crops to the egg parasitoid *Telenomus podisi* (Hymenoptera: Platygastridae). Neotrop Entomol 47:281-291. https://doi.org/10.1007/s13744-017-0552-9

Steinhaus EA, Warpechowski LF, Braga LE, Morin M, Tenório C, Boff JS., Bernardi O, Farias JR (2022) Intra-and interspecific variation in the susceptibility to insecticides of stink bugs (Hemiptera: Pentatomidae) that attack soybean and maize in southern Brazil. J Econ Entomol 115:631-636. https://doi.org/10.1093/jee/toac013

Therneau TM, Grambsch PM (2000) Modeling survival data: extending the Cox model. Springer, New York. https://doi.org/10.1007/978-1-4757-3294-8

Tessmer MA, Kuhn TMA, Appezzato-da-Glória B, Lopes JRS, Erler G, Bonani JP (2022) Histology of damage caused by *Euschistus heros* (F.) nymphs in soybean pods and seeds. Neotrop Entomol 51:112-121. https://doi.org/10.1007/s13744-021-00931-w

Theenoor R, Ghosh A, Venkatesan R (2024) Harmonizing Control: Understanding the complex impact of pesticides on parasitoid wasps for enhanced pest management. Curr Opin Insect Sci 65:101236. https://doi.org/10.1016/j.cois.2024.101236

Tibola CM, Silva L, Sgubin F, Omoto C (2021) Monitoring resistance of *Euschistus heros* (Fabricius) (Hemiptera: Pentatomidae) to insecticides by using encapsulated artificial diet bioassay. Insects 12:599. https://doi.org/10.3390/insects12070599

van Buuren S, Fredriks M (2001) Worm plot: a simple diagnostic device for modelling growth reference curves. *Stat Med* 20:1259–1277. https://doi.org/10.1002/sim.746

Van Lenteren JC, Bueno VH, Klapwijk JN (2021) Augmentative biological Control In: Mason PG (ed) Biological Control: Global impacts, challenges and future directions of pest management, 1st ed. Csiro Publishing, Melbourne, pp 90-109.

Vezzani FM, Anghinoni I, Cherubin MR, Mendes IC (2024) Soil health and modern Brazilian agriculture In: Karlen DL (ed) Soil Health Series: Volume 3 Soil Health and Sustainable Agriculture in Brazil, 1-18.

Weber ID, Garcia AG, Bueno AF, Oliveira RC, Godoy WAC (2022). Release strategies of *Telenomus podisi* for control of *Euschistus heros*: a computational modeling approach. Pest Manag Sci, 78 :4544-4556. https://doi.org/10.1002/ps.7074

Yadav P, Borges RM (2024). Oviposition decisions under environment-induced physiological stress in parasitoids. Curr Opin Insect Sci, 65:101240. https://doi.org/10.1016/j.cois.2024.101240

Zantedeschi R, Grützmacher AD, Pazini JDB, Bueno FA, Machado LL (2018a) Selectivity of pesticides registered for soybean crop on *Telenomus podisi* and *Trissolcus basalis*. Pesq Agropec Trop 48:52-58. https://doi.org/10.1590/1983-40632018v4850348

Zantedeschi R, Rakes M, Pasini RA, Araújo MB, Bueno FA, Grützmacher AD (2018b) Toxicity of soybean-registered agrochemicals to *Telenomus podisi* and *Trissolcus basalis* immature stages. Phytoparasitica 46:203-212. https://doi.org/10.1007/s12600-018-0660-z

Zhang Y, Lorsbach BA, Castetter S, Lambert WT, Kister J, Wang NX, Klittich CJ, Roth J, Sparks TC, Loso MR. Physicochemical property guidelines for modern agrochemicals (2018) Pest Manag Sci 74:1979-91. https://doi.org/10.1002/ps.5037

**Table 1.** Pesticides tested for their lethal and sublethal effects in *Telenomus podisi.*

| Active ingredient | Trade name | Formulation | Tested dose [1] (mL or g $ha^{-1}$) | Manufacturer |
|---|---|---|---|---|
| *Inseticidas* | | | | |
| Bifenthrin + carbosulfan | Talisman® | Emulsifiable concentrate (EC) | 600* | FMC Química do Brasil Ltda. |
| Acephate | Perito 970 SG® | Water-soluble granules (SG) | 1000* | UPL do Brasil Indústria e Comércio de Insumos Agropecuários S.A |
| Isocycloseram + lambda-cyhalothrin | Verdavis® | Suspension concentrate (SC) + Capsule suspension (CS) | 250* | Syngenta Proteção de cultivos Ltda. |
| Etiprole | Curbix 200 SC® | Suspension concentrate (SC) | 1000* | Bayer S.A. |
| *Fungicidas* | | | | |
| Bixafen + prothioconazole + trifloxystrobin | Fox® Xpro | Suspension concentrate (SC) | 500 | Bayer S.A. |
| Fenpropimorph | Versatilis® | Emulsifiable concentrate (CE) | 500 | BASF SE |
| Chlorothalonil | Bravonil 720® | Suspension concentrate (SC) | 2000 | Syngenta Proteção de cultivos Ltda. |
| *Herbicidas* | | | | |
| Glyphosate (SL formulation) | Roundup Original Mais® | Soluble concentrate (SL) | 6000 | Bayer S.A. |
| Glyphosate (WG formulation) | Roundup WG® | Water-dispersible granules (WG) | 3500 | Bayer S.A. |

*Rates correspond to the highest label-recommended doses used in pupal-exposure bioassays. For adult-exposure bioassays, insecticides were applied at their estimated $LC_{50}$ values (Table 2), whereas fungicides and herbicides were applied at the rates shown..

**Table 2.** Lethal concentration ($LC_{50}$ in mg a.i. $L^{-1}$) and respective confidence intervals of different insecticides on *Telenomus podisi* adults using topical spray bioassays.

| Treatments | n [1] | Slope (± SE) | $\chi^{2\,(2)}$ | df [3] | h [4] | $LC_{50}$ (IC) [5] |
|---|---|---|---|---|---|---|
| Bifenthrin + carbosulfan | 275 | 3.24 ± 0.29 | 1.26 | 2 | 0.63 | 42.97 (36.99 – 48.98) |
| Acephate | 290 | 3.79 ± 0.29 | 2.22 | 2 | 1.10 | 265.88 (200.90 – 346.56) |
| Isocycloseram + lambda-cyhalothrin | 649 | 1.70 ± 0.15 | 1.98 | 2 | 0.99 | 28.26 (23.32 – 34.46) |
| Etiprole | 672 | 1.58 ± 0.16 | 2.21 | 2 | 1.11 | 68.27 (41.51 – 122.16) |

[1]n: number of insects tested;
[2] $\chi$2: calculated chi-square value;
[3] d.f.: degrees of freedom;
[4] h: heterogeneity factor;
[5] IC: confidence interval of 95%;

**Table 3.** Cumulative mortality of adults and parasitism rate of *Telenomus podisi*, in the $F_0$ generation, when adults were exposed to insecticides and their binary mixtures.

| Treatments | Mortality[1] (%) | Parasitism (%)[2] |
|---|---|---|
| Control | 2.50 ± 2.50 j | 89.50 ± 3.43 a |
| Isocycloseram + lambda-cyhalothrin | 54.60 ± 3.47 ghi | 48.10 ± 0.74 efg |
| Etiprole | 47.92 ± 4.58 hi | 50.73 ± 3.99 defg |
| Acephate | 45.48 ± 4.43 hi | 53.29 ± 12.63 defg |
| Bifenthrin + carbosulfan | 46.70 ± 5.31 i | 66.98 ± 1.58 bcde |
| Bixafen + prothioconazole + trifloxystrobin | 7.50 ± 4.79 j | 79.51 ± 3.32 ab |
| Fenpropimorph | 4.58 ± 2.67 j | 78.43 ± 3.55 abc |
| Chlorothalonil | 6.79 ± 4.72 j | 73.56 ± 4.34 abcd |
| Glyphosate (SL formulation) | 8.12 ± 4.75 j | 66.06 ± 4.29 bcde |
| Glyphosate (WG formulation) | 5.05 ± 2.94 j | 70.49 ± 3.69 bcde |
| Isocycloseram + lambda-cyhalothrin + Bixafen + prothioconazole + trifloxystrobin | 72.41 ± 9.57 fghi | 32.46 ± 13.86 h |
| Isocycloseram + lambda-cyhalothrin + Fenpropimorph | 76.50 ± 4.32 def | 34.80 ± 11.00 gh |
| Isocycloseram + lambda-cyhalothrin + Chlorothalonil | 82.92 ± 7.47 bcdef | 10.45 ± 6.69 i |
| Isocycloseram + lambda-cyhalothrin + Glyphosate (SL formulation) | 92.73 ± 2.43 abc | 2.42 ± 2.42 i |
| Isocycloseram + lambda-cyhalothrin + Glyphosate (WG formulation) | 96.88 ± 3.12 ab | -[#] |
| Etiprole + Bixafen + prothioconazole + trifloxystrobin | 74.23 ± 2.52 def | 54.80 ± 4.96 def |
| Etiprole + Fenpropimorph | 85.42 ± 3.56 bcde | 54.42 ± 14.57 defg |
| Etiprole + Chlorothalonil | 78.41 ± 2.84 cdef | 55.95 ± 7.27 cdef |
| Etiprole + Glyphosate (SL formulation) | 100.00 ± 0.00* | -[#] |
| Etiprole + Glyphosate (WG formulation) | 93.51 ± 2.33 ab | -[#] |
| Acephate + Bixafen + prothioconazole + trifloxystrobin | 66.02 ± 6.70 fgh | 30.48 ± 5.96 gh |
| Acephate + Fenpropimorph | 97.73 ± 2.27 a | 55.26 ± 3.62 def |
| Acephate + Chlorothalonil | 70.00 ± 7.07 efg | 51.18 ± 6.25 defg |
| Acephate + Glyphosate (SL formulation) | 94.64 ± 3.42 ab | 47.31 ± 4.71 efg |
| Acephate + Glyphosate (WG formulation) | 75.43 ± 5.32 def | 56.02 ± 20.57 bcdef |
| Bifenthrin + carbosulfan + Bixafen + prothioconazole + trifloxystrobin | 79.17 ± 7.50 cdef | 65.00 ± 8.38 bcde |
| Bifenthrin + carbosulfan + Fenpropimorph | 90.87 ± 5.51 abcd | 54.43 ± 12.16 defg |
| Bifenthrin + carbosulfan + Chlorothalonil | 77.64 ± 9.13 def | 32.47 ± 1.62 fgh |
| Bifenthrin + carbosulfan + Glyphosate (SL formulation) | 92.92 ± 2.39 ab | 51.99 ± 4.78 defg |
| Bifenthrin + carbosulfan + Glyphosate (WG formulation) | 87.35 ± 2.82 bcd | 63.30 ± 2.59 bcde |
| $\chi^2$ | 620.26 | 132.82 |
| Df | 28 | 26 |
| *P* | < 0.01 | < 0.01 |

Means ± standard error followed by the same letter in a column do not differ significantly according to pairwise likelihood-ratio tests (p > 0.05). [1]Binomial GLM.

[2]Beta-binomial GLM.
*Treatment not included in the statistical analysis (zero variance).
#Not evaluated due to the high mortality rate.

**Table 4.** Emergence and sex ratio of *Telenomus podisi*, $F_1$ generation, when adults were exposed to insecticides and their binary mixtures.

| **Treatments** | **Emergence (%)[1]** | **Sex ratio (%)[2#]** |
|---|---|---|
| Control | 88.71 ± 1.76 a | 89.50 ± 2.15 a |
| Isocycloseram + lambda-cyhalothrin | 32.24 ± 6.28 gh | 83.03 ± 2.25 abcd |
| Etiprole | 77.85 ± 4.40 bc | 75.43 ± 4.03 bcde |
| Acephate | 66.31 ± 8.68 bcd | 61.91 ± 16.09 ef |
| Bifenthrin + carbosulfan | 66.00 ± 8.60 cd | 82.34 ± 1.72 abcd |
| Bixafen + prothioconazole + trifloxystrobin | 72.17 ± 4.67 bcd | 76.60 ± 2.54 bcde |
| Fenpropimorph | 80.93 ± 3.64 ab | 79.67 ± 2.49 abcde |
| Chlorothalonil | 67.57 ± 6.47 cd | 69.54 ± 2.71 def |
| Glyphosate (SL formulation) | 67.83 ± 4.31 cd | 70.79 ± 2.91 cdef |
| Glyphosate (WG formulation) | 74.26 ± 1.55 bcd | 80.23 ± 2.18 abcd |
| Isocycloseram + lambda-cyhalothrin + Bixafen + prothioconazole + trifloxys-trobin | 28.76 ± 9.00 h | 83.57 ± 5.31 abcd |
| Isocycloseram + lambda-cyhalothrin + Fenpropimorph | 66.39 ± 8.24 cde | 82.41 ± 5.09 abcde |
| Isocycloseram + lambda-cyhalothrin + Chlorothalonil | 63.33 ± 3.33 bcdef | 75.71 ± 4.29 abcdef |
| Etiprole + Bixafen + prothioconazole + trifloxystrobin | 70.60 ± 3.08 bcd | 83.57 ± 2.78 ab |
| Etiprole + Fenpropimorph | 61.81 ± 4.95 de | 82.74 ± 3.13 abcd |
| Etiprole + Chlorothalonil | 50.51 ± 4.84 ef | 83.49 ± 4.13 ab |
| Acephate + Bixafen + prothioconazole + trifloxystrobin | 39.14 ± 7.55 fgh | 84.42 ± 4.55 ab |
| Acephate + Fenpropimorph | 68.02 ± 2.19 cd | 83.82 ± 2.93 ab |
| Acephate + Chlorothalonil | 40.72 ± 8.18 fgh | 62.49 ± 2.81 f |
| Acephate + Glyphosate (SL formulation) | 78.19 ± 5.69 abc | 45.16 ± 4.38 g |
| Acephate + Glyphosate (WG formulation) | 44.44 ± 18.23 bcd | 72.12 ± 3.36 bcdef |
| Bifenthrin + carbosulfan + Bixafen + prothioconazole + trifloxystrobin | 76.57 ± 2.78 bc | 83.82 ± 4.95 ab |
| Bifenthrin + carbosulfan + Fenpropimorph | 78.10 ± 7.53 bcd | 67.33 ± 17.22 abc |
| Bifenthrin + carbosulfan + Chlorothalonil | 45.66 ± 6.42 efg | 77.42 ± 6.72 abcde |
| Bifenthrin + carbosulfan + Glyphosate (SL formulation) | 68.44 ± 3.59 cd | 41.61 ± 10.90 g |
| Bifenthrin + carbosulfan + Glyphosate (WG formulation) | 71.51 ± 7.08 bcd | 77.84 ± 3.06 abcde |
| $\chi^2$ | 201.14 | 82.70 |
| Df | 25 | 25 |
| *P* | < 0.01 | < 0.01 |

Means ± standard error followed by the same letter in a column do not differ significantly according to pairwise likelihood-ratio tests (p > 0.05). [1]Binomial GLM.
[2]Beta-binomial GLM.
#Percentage of emerged females.

**Table 5.** Emergence rate and parasitism of *Telenomus podisi* in the $F_0$ generation when pupae were exposed to insecticides and their binary mixtures.

| **Treatments** | **Emergence (%)**[1] | **Parasitism (%)**[2] |
|---|---|---|
| Control | 81.71 ± 2.94 a | 92.79 ± 2.23 a |
| Isocycloseram + lambda-cyhalothrin | 0.00 ± 0.00* | -# |
| Etiprole | 67.43 ± 2.65 bc | 89.43 ± 1.58 ab |
| Acephate | 17.14 ± 2.02 f | 77.86 ± 11.88 bcd |
| Bifenthrin + carbosulfan | 0.00 ± 0.00* | -# |
| Bixafen + prothioconazole + trifloxystrobin | 81.71 ± 2.49 a | 90.43 ± 2.45 ab |
| Fenpropimorph | 34.86 ± 10.12 e | 88.08 ± 3.94 abc |
| Chlorothalonil | 82.86 ± 5.50 a | 85.47 ± 2.84 abc |
| Glyphosate (SL formulation) | 59.43 ± 6.22 cd | 85.97 ± 2.60 abc |
| Glyphosate (WG formulation) | 64.57 ± 3.08 bc | 86.50 ± 1.78 abc |
| Isocycloseram + lambda-cyhalothrin + Bixafen + prothioconazole + trifloxystrobin | 0.00 ± 0.00* | -# |
| Isocycloseram + lambda-cyhalothrin + Fenpropimorph | 0.57 ± 0.57 h | -# |
| Isocycloseram + lambda-cyhalothrin + Chlorothalonil | 0.00 ± 0.00* | -# |
| Isocycloseram + lambda-cyhalothrin + Glyphosate (SL formulation) | 0.00 ± 0.00* | -# |
| Isocycloseram + lambda-cyhalothrin + Glyphosate (WG formulation) | 0.00 ± 0.00* | -# |
| Etiprole + Bixafen + prothioconazole + trifloxystrobin | 74.29 ± 4.69 ab | 72.46 ± 18.19 d |
| Etiprole + Fenpropimorph | 48.57 ± 4.24 d | 53.49 ± 17.38 d |
| Etiprole + Chlorothalonil | 64.57 ± 4.48 bc | 68.84 ± 17.33 d |
| Etiprole + Glyphosate (SL formulation) | 0.00 ± 0.00* | -# |
| Etiprole + Glyphosate (WG formulation) | 0.00 ± 0.00* | -# |
| Acephate + Bixafen + prothioconazole + trifloxystrobin | 10.29 ± 1.46 fg | 52.42 ± 5.43 d |
| Acephate + Fenpropimorph | 5.14 ± 2.46 g | -# |
| Acephate + Chlorothalonil | 14.29 ± 3.38 f | 73.74 ± 9.01 cd |
| Acephate + Glyphosate (SL formulation) | 0.00 ± 0.00* | -# |
| Acephate + Glyphosate (WG formulation) | 0.00 ± 0.00* | -# |
| Bifenthrin + carbosulfan + Bixafen + prothioconazole + trifloxystrobin | 0.00 ± 0.00* | -# |
| Bifenthrin + carbosulfan + Fenpropimorph | 0.00 ± 0.00* | -# |
| Bifenthrin + carbosulfan + Chlorothalonil | 0.00 ± 0.00* | -# |
| Bifenthrin + carbosulfan + Glyphosate (SL formulation) | 0.00 ± 0.00* | -# |
| Bifenthrin + carbosulfan + Glyphosate (WG formulation) | 0.00 ± 0.00* | -# |
| $\chi^2$ | 180.47 | 50.40 |
| Df | 14 | 12 |
| *P* | < 0.01 | < 0.01 |

Means ± standard error followed by the same letter in a column do not differ significantly according to pairwise likelihood-ratio tests (p > 0.05). [1]Beta-binomial GLM. [2]Beta-binomial GAMLSS with the effect of treatments on the mean and dispersion.
*Treatment not included in the statistical analysis (zero variance).
#Not evaluated due to high mortality rate.

**Table 6.** Emergence and sex ratio of *Telenomus podisi* in the $F_1$ generation when pupae were exposed to insecticides and their binary mixtures.

| **Treatments** | **Emergence (%)**[1] | **Sex ratio (%)**[1#] |
|---|---|---|
| Control | 86.37 ± 1.87 a | 85.71 ± 4.54 a |
| Etiprole | 69.48 ± 4.32 cd | 72.81 ± 16.56 a |
| Acephate | 59.13 ± 6.83 de | 30.00 ± 16.77 bc |
| Bixafen + prothioconazole + trifloxystrobin | 83.28 ± 3.57 abc | 81.20 ± 4.48 a |
| Fenpropimorph | 85.37 ± 3.50 ab | 84.25 ± 1.76 a |
| Chlorothalonil | 70.94 ± 6.02 cd | 81.75 ± 3.66 a |
| Glyphosate (SL formulation) | 81.23 ± 3.64 abc | 74.20 ± 3.83 a |
| Glyphosate (WG formulation) | 81.47 ± 3.50 abc | 78.80 ± 3.50 a |
| Etiprole + Bixafen + prothioconazole + trifloxystrobin | 76.91 ± 4.72 abcd | 63.87 ± 21.82 ab |
| Etiprole + Fenpropimorph | 47.62 ± 17.71 e | 22.22 ± 22.22 cd |
| Etiprole + Chlorothalonil | 70.05 ± 3.72 cd | 83.73 ± 2.78 a |
| Acephate + Bixafen + prothioconazole + trifloxystrobin | 70.35 ± 8.09 bcd | 4.32 ± 2.70 d |
| Acephate + Chlorothalonil | 73.72 ± 4.39 abcd | 21.94 ± 15.62 cd |
| $\chi^2$ | 31.58 | 56.92 |
| Df | 12 | 12 |
| *P* | < 0.01 | < 0.01 |

Means ± standard error followed by the same letter in a column do not differ significantly according to pairwise likelihood-ratio tests (p > 0.05). [1]Beta-binomial GLM.
[#]Percentage of emerged females.

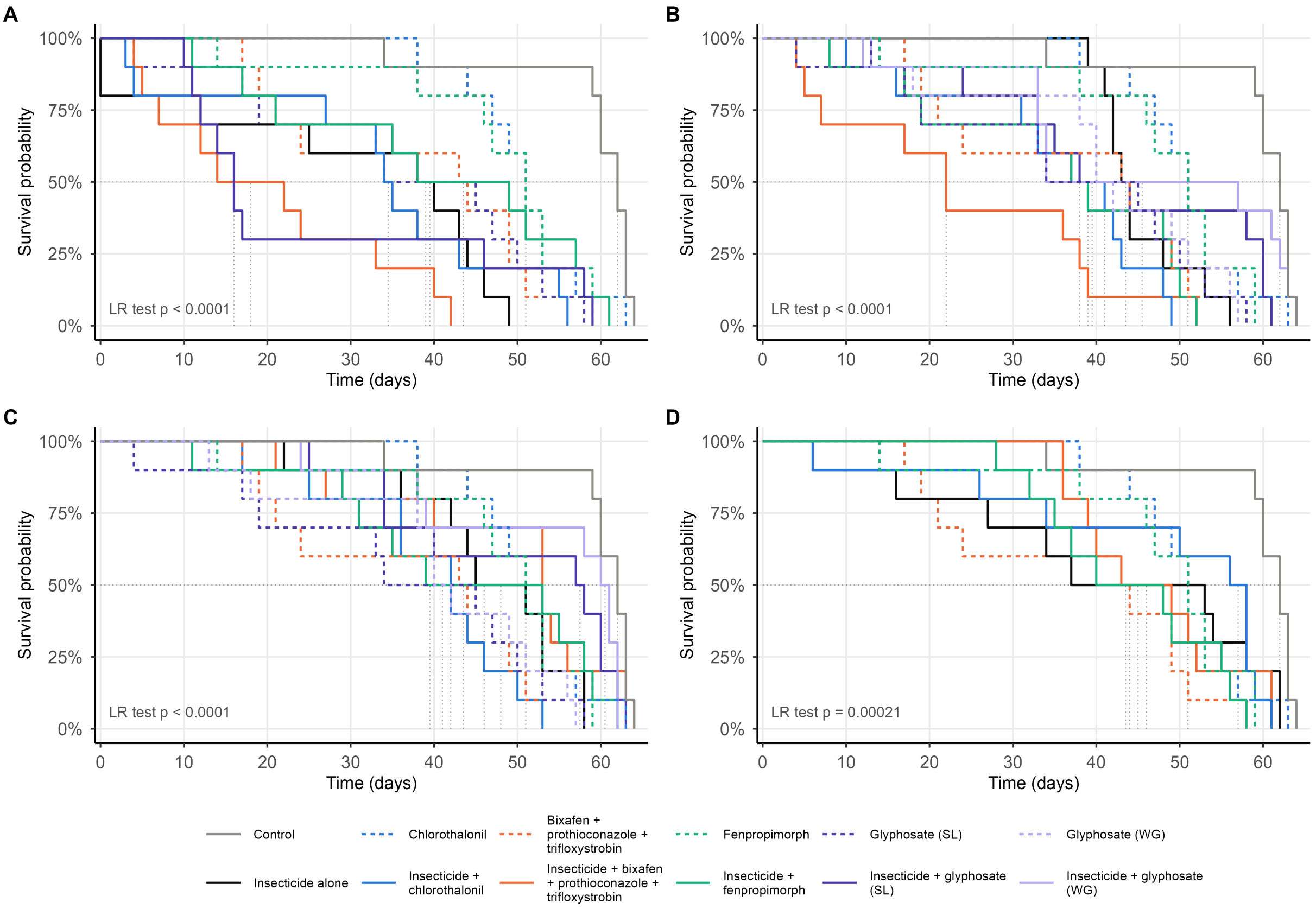

**Fig. 1.** Survival probability of adult *Telenomus podisi* over time following exposure to the LC50 of isocycloseram + lambda-cyhalothrin (A), acephate (B), bifenthrin + carbosulfan (C), and ethiprole (D), applied either alone or in binary mixtures with one of the tested synthetic fungicides or herbicides. Colors identify the fungicide or herbicide; dashed lines represent the product applied alone and solid lines its binary mixture with the insecticide of each panel. Gray and black lines represent the negative control and the insecticide applied alone, respectively. Dotted lines indicate the median survival time for each treatment, and p-values are from likelihood-ratio tests of the Cox proportional-hazards models.

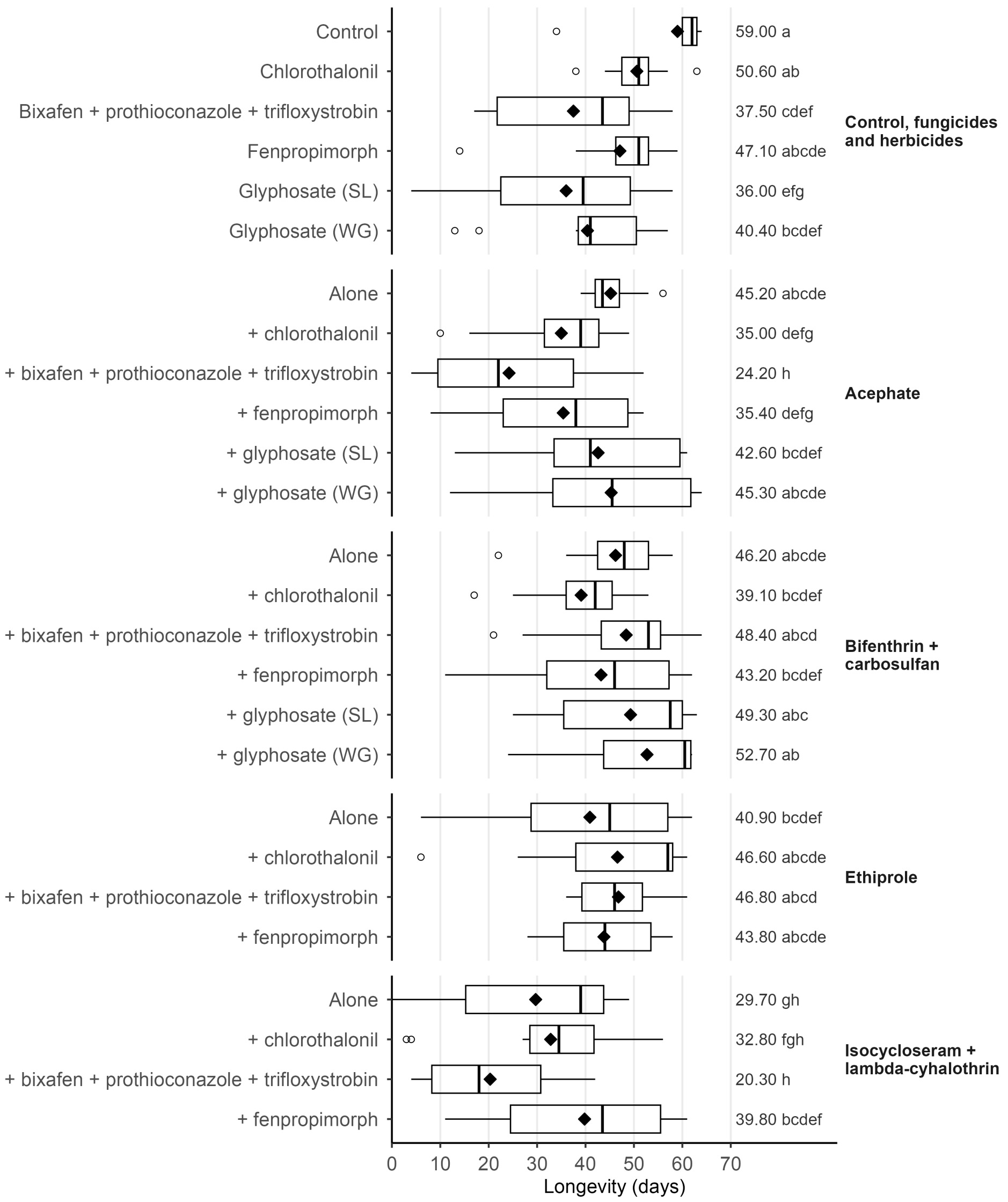

**Fig. 2.** Longevity of adult *Telenomus podisi* ($F_0$ generation) exposed to insecticides at their respective $LC_{50}$ values and to fungicides and herbicides at their label-recommended rates, applied either individually or in binary mixtures. Treatments are grouped by insecticide ("Alone", insecticide applied alone; "+", binary mixture with the indicated fungicide or herbicide). Vertical lines within the boxes represent the median, boxes extend from the first to the third quartile (interquartile range, IQR), whiskers extend to the lowest and highest values within 1.5 × IQR, open circles indicate outliers, and diamonds represent the mean. Values on the right are means followed by letters; treatments sharing a letter do not differ significantly according to pairwise likelihood-ratio tests of the negative binomial type II GLM ($p > 0.05$).

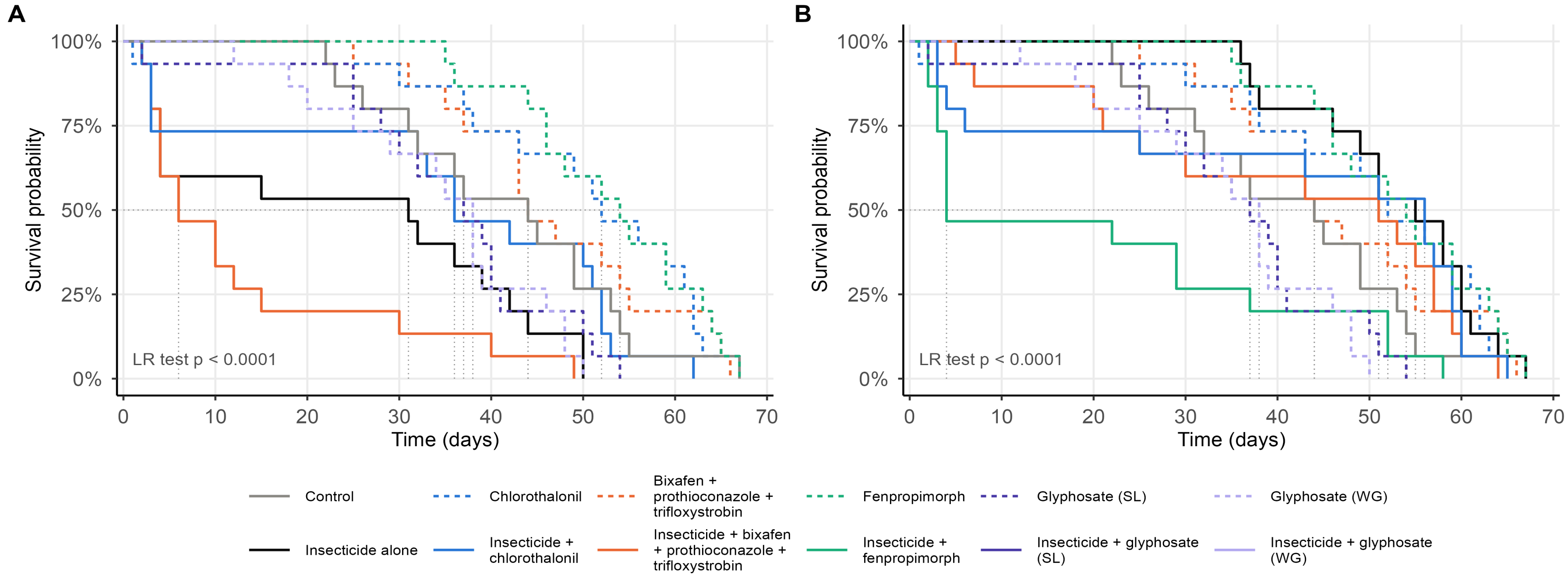

**Fig. 3.** Survival probability of adult *Telenomus podisi* following exposure of the parasitoid pupal stage to acephate (A) and ethiprole (B), applied either alone or in binary mixtures with one of the tested synthetic fungicides or herbicides. Colors and line types as in Fig. 1. Dotted lines indicate the median survival time for each treatment, and p-values are from likelihood-ratio tests of the Cox proportional-hazards models.

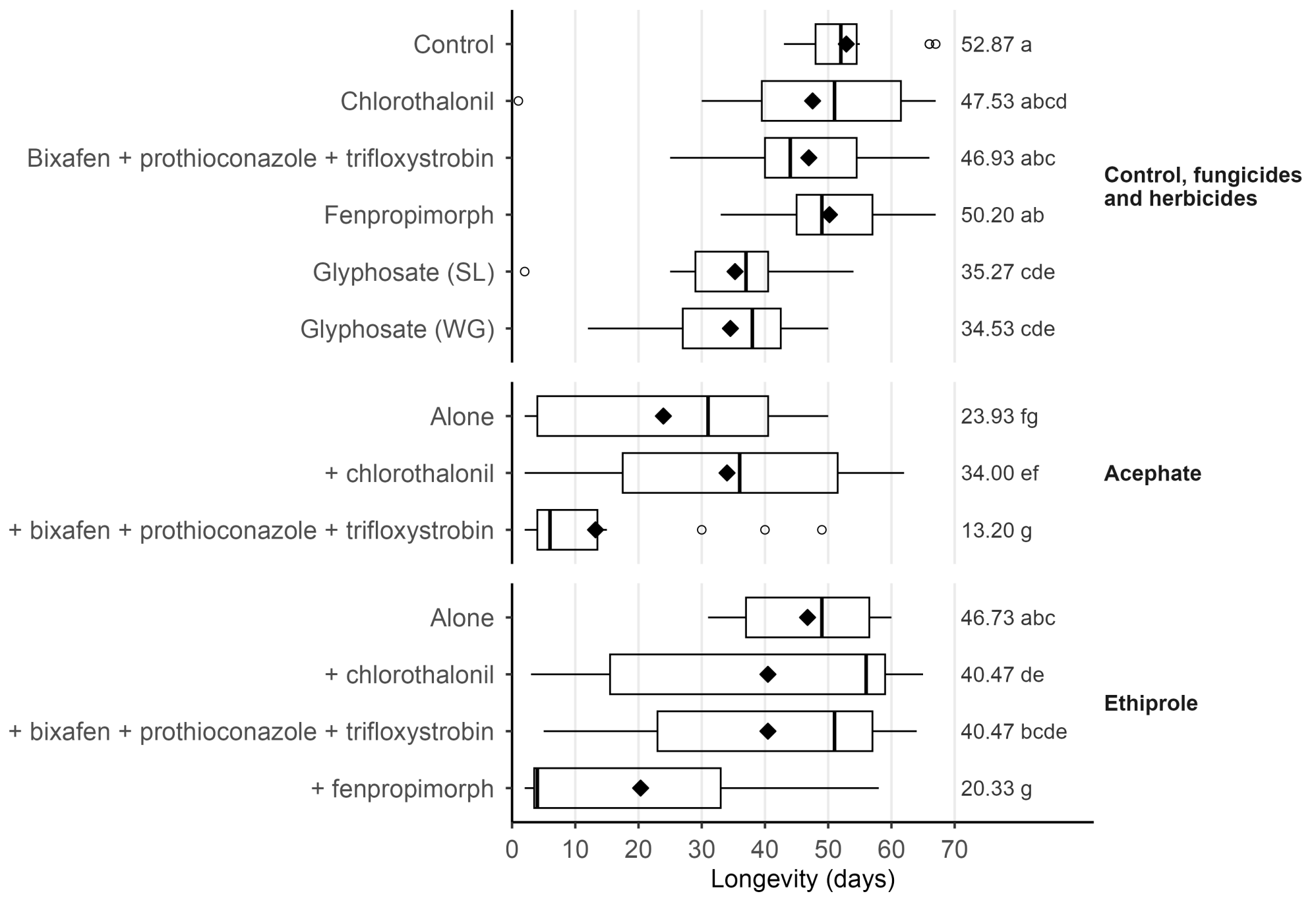

**Fig. 4.** Longevity of adult *Telenomus podisi* ($F_0$ generation) following exposure of the pupal stage to the tested insecticides, fungicides, and herbicides, applied either individually or in binary mixtures. Treatments, boxplots, and letters as in Fig. 2.

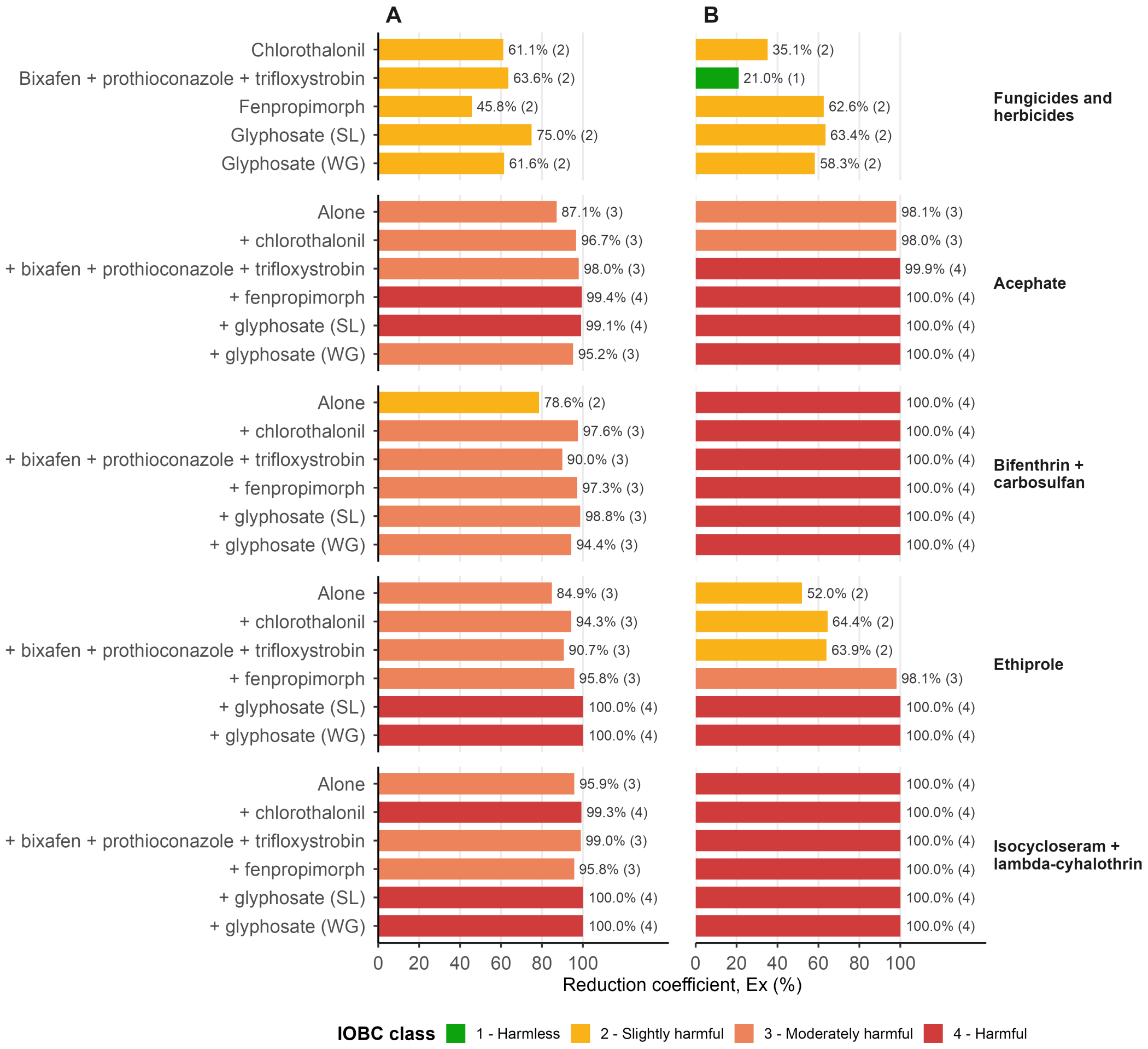

**Fig. 5.** Reduction coefficient (Ex) of pesticide treatments and their classification according to IOBC/WPRS toxicity classes for *Telenomus podisi* after exposure during the adult stage (A) and pupal stage (B). Treatments are grouped by insecticide as in Fig. 2. Bars represent Ex (%), and colors indicate the IOBC/WPRS classes, which are also given in brackets after each value: 1 = harmless (Ex < 30%), 2 = slightly harmful (30% ≤Ex < 80%), 3 = moderately harmful (80% ≤Ex ≤99%), and 4 = harmful (Ex > 99%).